\documentclass[a4paper,10pt]{article}
\usepackage{epsf,latexsym,amsmath}
\usepackage{times}
\usepackage{amssymb}
\usepackage{graphicx}

\begin{document}
\begin{center}

{\Large Efficient quantum processing \\
 of 3--manifold topological  invariants}
 
\vskip .5cm

{\large S Garnerone$^{\dag \, \sharp\, (1)}$,
 A Marzuoli$^{\ddagger \,\sharp\,(2)}$ 
and M Rasetti$^{\dag \, \sharp\,(3)}$}

\vskip 1cm

{\small $^{\dag}$Dipartimento di Fisica, Politecnico di Torino, corso Duca degli Abruzzi 24, 10129 Torino (Italy);\\
$^{\sharp}$ Institute for Scientific Interchange, Villa Gualino, Viale Settimio Severo 75, 10131 Torino (Italy);\\
$^{\ddagger}$Dipartimento di Fisica Nucleare e Teorica, Universit\`a degli Studi di Pavia and Istituto 
Nazionale di Fisica Nucleare, Sezione di Pavia, via A. Bassi 6, 27100 Pavia (Italy)}
\end{center}

\vskip 1cm

\begin{abstract}
A quantum algorithm for approximating efficiently 
$3$--manifold topological  invariants in the framework of
$SU(2)$ Chern--Simons--Witten (CSW) topological  quantum field theory
at finite values of the coupling constant $k$
is provided.
The model of computation adopted is the $q$--deformed spin network model
viewed as a quantum recognizer in the sense of \cite{WiCr},
where each basic unitary transition function can
be efficiently processed by a standard quantum circuit.

This achievement is an extension of the algorithm for approximating polynomial
invariants of colored oriented links found in \cite{GaMaRa1,GaMaRa2}. 
Thus all the significant 
quantities --partition functions and
 observables-- of quantum CSW theory can be processed efficiently on a quantum computer,
 reflecting the intrinsic, field--theoretic solvability of such theory at finite $k$.
 
 The paper is supplemented by a critical overview of the
basic conceptual tools underlying the construction of quantum 
invariants of links and $3$--manifolds and connections with
algorithmic questions that arise in geometry and quantum gravity models are discussed.
\end{abstract}

\vspace{30pt}


\vspace{5pt}


\vspace{70pt}

\noindent
---------------------------------------\\
(1) silvano.garnerone@polito.it\\
(2) annalisa.marzuoli@pv.infn.it\\
(3) mario.rasetti@polito.it\\

\vfill
\newpage

\section{Introduction}

The possibility of computing quantities of topological or geometric nature
was recognized as a major achievement for quantum information theory
in a series of paper  by Michael Freedman and co--workers 
\cite{Fre,FrKiWa,FrLaWa}
(see also \cite{Pre} for an introduction). 
Their `topological quantum computation' setting,
intrinsically fault--tolerant and protected
from decoherence, was
designed to comply with the behavior of `modular functors' of 
$3D$ Chern--Simons--Witten (CSW) non--abelian topological quantum
field theory (TQFT) \cite{Wit,Ati,BiBlRa}, the gauge group being typically $SU(2)$. 
In physicists' language, such functors are partition functions 
and correlators of the quantum theory
and, owing to gauge invariance and invariance under
diffeomorphisms, which freeze out local degrees of freedom, 
they share a global, `topological' character.
More precisely, the physical observables are associated with
topological invariants of knots --the prototype of which 
is the Jones polynomial \cite{Jon}-- and the generating functional
is an invariant of the 
$3$--dimensional ambient manifold, the 
Reshetikhin--Turaev--Witten invariant \cite{ReTu,Wit}.

The search for efficient quantum algorithms to compute
(approximations of) knot invariants has been carried out
by several groups in the last few years.\\
Within the framework of topological
quantum computation the existence of an
effective procedure has been taken for granted
by resorting to the fact that this model 
is polynomially reducible to
the standard quantum circuit model \cite{NiCh}.  
Later on, this implicit proof has been supported by the fundamental notion 
of `additive approximation' introduced in \cite{BoFrLo} and 
borrowed in all the alternative approaches.\\
The first explicit algorithm for the Jones polynomial
of a knot presented as the plat closure of a braid
has been constructed in \cite{AhJoLa} by efficiently approximating 
unitary matrices associated with a particular representation 
of the braid group,
while in \cite{WoYa} the knot presentation 
and the representation of the braid group
are generalized (these notions will be  
defined in the following sections). 
It is worth stressing that both these approaches rely on the standard
model of quantum computation, namely the quantum
circuit model based on qubits, elementary
unitary gates and related algorithmic techniques
such as the so--called Hadamard trick \cite{NiCh}.\\
The approach we proposed in \cite{GaMaRa1,GaMaRa2}
differs from the previous ones in many respects.
First, we were able to handle more general knot invariants,
namely `colored' link polynomials (a link is a multicomponent knot) 
which reduces to Jones' in a particular case. Secondly, the
model of quantum computation we adopted, that we refer to as the $q$--deformed
spin network model, is in some sense a blending 
of the computational schemes mentioned above framed
within the background provided by the theory of
finite--states quantum machines \cite{WiCr}. 
This topic will be addressed in Section 2, while
a few conceptual questions and implications of our model of computation
will be discussed in the last section of this introduction.

In the main part of the paper, section 3,
we extend our efficient algorithm for 
approximating colored link polynomial to deal with 
`quantum' $3$--manifold invariants.

In order to overcome the difficulties 
due to the fact that we shall resort to concepts 
and definitions arising in many different contexts (low--dimensional
topology, quantum group theory, CSW field theory, $3$--dimensional
quantum gravity models,
classical and quantum complexity theory), we shall illustrate in the next 
few paragraphs the basic conceptual tools
underlying the construction of such `universal' invariants.

\subsection{From quantum topology to topological quantum field theory}

The term `quantum topology' was introduced by Turaev \cite{Tur}
to denote implications on the topological side of the algebraic
theory of quantum groups --technically, deformations of universal
enveloping algebras of Lie groups. The latter, based on the pioneering work
of Drinfel'd \cite{Dri} and Jimbo \cite{Jim}, was inspired by theoretical physics
 from its very beginning since quantum groups and associated $R$--matrix 
representations are the basic tools of quantum inverse scattering methods
 and are the backbone of exactly solvable models  in statistical mechanics \cite{Wu}.
 
 The deformation parameter $q$ was originally assumed to be a real number 
 related to Planck constant by $q= e^{h}$, therefore  it is commonly referred
 to as a `quantum' deformation, while the `classical', undeformed Lie group simmetry is
 recovered at the particular value $q=1$ ($h \rightarrow 0)$.
 When dealing with quantum invariants of knots and $3$--manifolds
 \cite{Tur,Oht} $q$ is most often a complex root of unity, the case $q=1$
 being considered as the `trivial' one. However, in a topologist's language,
 `classical' topological invariants are not the $(q=1)$--counterparts of
 `quantum' invariants, but rather the usual invariants of
 algebraic and geometric topology, typically related to the fundamental
 group and homology groups of manifolds and submanifolds.
 
 As Roberts remarks in the introduction to \cite{Oht}
 the standard topological invariants were {\em created} in order to distinguish
 between things and, owing to their intrinsic definitions, it is clear what
 kind of properties they reflect.
 For instance, the Euler number $\chi$ of a smooth, closed and oriented
 surface ${\cal S}$ determines completely its topological type and can be defined
 as $\chi ({\cal S}) =2-2g$, where $g$ is the number of handles of ${\cal S}$.
 On the other hand, quantum invariants of knots and $3$--manifolds
 were {\em discovered}, but their indirect constuction based on 
quantum group technology often hides information about the purely topological properties
they are able to detect.

 What is lost at the topological level is however well paid back by
the possibility of bridging this theory with a plenty of issues in pure mathematics
and theoretical physics ({\em cfr.} the review \cite{Oht} and the list of references therein). 
To  the early connections mentioned above (quantum inverse scattering, 
exact solvable models) it is worth 
adding the operator algebra approach used originally by Jones 
in defining his knot polynomial \cite{Jon}. However, the most profitable development
of the theory  was that suggested by Schwarz and formalized by Witten \cite{Wit}
(see \cite{KaGoRa} for a review and original references).

Indeed, recognizing quantum invariants as partition functions and vacuum 
expectation values of physical observables in  
Chern--Simons--Witten 
topological quantum field theory provides a `physical' explanation of 
their existence and properties. Even more radically, one could speak  of a `conceptual'
explanation, as far as the topological origin of these invariants keeps on
being unknown. In this wider sense, quantum topology might be thought of
as the mathematical substratum of an $SU(2)$ CSW topological field theory 
quantized according to the path integral prescription (the coupling constant $k \geq 1$
is constrained to be an integer related to the deformation parameter $q$ by $q=\exp (\tfrac{2\pi i}{k+2})$).\\ 
The CSW environment provides not only the physical interpretation of quantum
invariants but it does include as well all the historically distinct definitions \cite{Gua}.
In particular, monodromy representations of the braid group \cite{Koh} appear in
a variety of conformal field theories since point--like `particles' confined in $2$--dimensional
regions evolve along braided worldlines (\cite{GoRuSi} and references therein).
As a matter of fact, the natural extension of CSW theory 
to a $3$--manifold ${\cal M}^3$
endowed with a non empty $2$--dimensional boundary $\partial {\cal M}^3$
induces on $\partial {\cal M}^3$ a specific quantized boundary conformal
field theory, namely the $SU(2)$
Wess--Zumino--Witten (WZW) theory at level $\ell =k+2$ \cite{Wit,Car1}. 
The latter provides in turn the framework for
dealing with $SU(2)_q$--colored links presented as closures of oriented
braids and associated with Kaul unitary representation of the braid
group \cite{Kau1,RaGoKa}. A further extension of this representation 
proposed by the same author in \cite{Kau2}
is used in this paper to construct explicitly the quantum $3$--manifold invariants
in the form originally defined in \cite{KiMe} within a purely algebraic setting.  
Such quantities are essentially the Reshetikhin--Turaev--Witten
invariants \cite{ReTu} evaluated for $3$--manifolds presented as complements of
knots/links in the $3$--sphere $S^3$, up to an overall normalization.\\

\subsection{Algorithmic complexity of Chern--Simons--Witten theory}

As mentioned in the introductory remarks, the `quantum field' computer \cite{Fre,FrKiWa,FrLaWa}
is a model of computation designed to process quantities of topological
nature arising  in CSW environment and thus  
`effienciency' of any calulation --such as that of Jones knot polynomial--
should be guaranteed by definition (we leave aside here the issue of `exact' 
{\em versus} `approximate' calculation at least for the moment).\\ 
However, when dealing with algorithmic questions, the
model of computation adopted should comply with the commonly accepted
paradigms of theoretical computer science. 
Turing machines, together with  the polynomially equivalent 
 circuit models based on elementary boolean
gates,  represent the  universal schemes which allow 
problems and algorithms to  be grouped into classical complexity classes \cite{GaJo}.
In quantum computing, the notions of quantum Turing machine --and associated quantum
circuits based on qubits and unitary elementary gates--
 can be introduced and represent the standard, universal model 
of computation \cite{NiCh}.\\
It was shown in \cite{FrKiWa} that $SU(2)$  CSW functors at the fifth root of
unity, whose domain is restricted to
 collections of `topological' qubits (disks with three marked points) 
on which suitable unitary representations  of the braid groups ${\bf B}_3$ and 
${\bf B}_3 \times {\bf B}_3$ act, do reproduce the standard elementary
gates of the  quantum circuit model. The physical support of such information processing
consists of anyonic systems obeying particular types of braid statistics, and
work is in progress on the experimental side to check the implementability
of such approach
(see \cite{FrNaSh} and also \cite{ZhScTe} for
a different kind of implementation of braiding operators).\\ 
Based the above properties, a sort of `minimal' realization
of the full quantum field computer is (polynomially equivalent to) the quantum circuit,
and indeed the work of \cite{AhJoLa} and \cite{WoYa} on quantum computation of the Jones invariant 
does not depend at all on  any field theoretic background. 
Generally speaking, this  
is satisfactory for the aim of introducing an {\em ad hoc}
computing model for treating  anyonic quantum systems; yet is somehow disturbing
because classical Turing machines and their probabilistic counterparts
are able to simulate efficiently only  any dynamical systems governed by classical 
laws at any degree of accuracy. 
Even the objection that we are in the presence of a quantum field theory
--not simply a quantum mechanical many--body system--
is misleading since the CSW model is exacly solvable at the full quantum level
(for each fixed value of the coupling constant) without resorting to
any approximation such as the low--energy limit \cite{KaGoRa,Gua}. 
The crucial feature of possessing
only global, purely topological degrees of freedom makes quantum CSW theory
likely to be simulated within a computational scheme based on a discrete
space of states and able to implement polylocal braiding
operations.\\
As will be illustrated in section 2, the universal model of computation able to handle
all discrete, many--body quantum systems described by  (real or virtual) 
pure angular momenta states (not simply
two--level systems) as well as solvable field theories
of CSW--type is the ($q$--deformed) spin network proposed in \cite{GaMaRa1,GaMaRa2}.

\vspace{12pt}

As already recognized in \cite{Fre} the task of computing the Jones polynomial of a knot or link 
represents a major achievement since it is the simplest observable in quantum CSW theory and
then its calculation is a testing ground of the effectiveness of topological quantum computation.
However, the interest in this algorithmic problem has recently grown   in connection with
the search for new testing grounds for quantum information theory in general, without necessarily
exploiting the physical meaning of the invariant.

The reason why Jones link polynomial is so
crucial in the computational context relies on the fact that a `simpler'
link invariant, the Alexander--Conway polynomial, can be computed efficiently, while
the problem of computing $2$--variable polynomials --such as the HOMFLY invariant--
is $\mathbf{NP}$--hard (see \cite{BiBr,Oht,JaVeWe} for definitions
of these invariants, original references and
accounts of algorithmic questions).

The issue of computational complexity of the $1$--variable Jones polynomial in classical information theory
can be summarized in
\begin{quote}
{\em \underline{Problem 1}\\ 
How hard is it to determine the Jones polynomial $\mathit{J} (L,q)$
 of a link $L$? }
\end{quote} 
A quite exhaustive answer has been provided in (JVW), where the evaluation of the Jones polynomial
of an alternating link $\tilde{L}$ at a root of unity $q$ is shown to be 
$\mathbf{\# P}$--hard, namely computationally intractable in a very strong sense.\\
 Recall first that
`alternating' links are special instances of links, the planar diagrams of which exhibit
over and under crossings, alternatively. Thus the evaluation of the invariant of generic,
not only alternating links is at least as hard. Secondly, the computation becomes feasible 
when the argument $q$ of the polynomial is a  2nd, 3rd, 4th, 6th  root of unity,
so that the first not easy case involves a 5th root of unity 
(refer to \cite{JaVeWe} for details on this technical issue). 
Finally, the $\mathbf{\# P}$ complexity class can be defined as the class of enumeration problems
in which the structures that must be counted are recognizable in polynomial time.
A problem $\pi$ in $\mathbf{\# P}$ is said to be $\mathbf{\# P}$--complete if for any other problem $\pi '$ in
$\mathbf{\# P}$, $\pi '$ is polynomial--time reducible to $\pi$; if a polynomial time algorithm were found
for any such problem, it would follow that $\mathbf{\# P} \subseteq \mathbf{P}$. A problem is $\mathbf{\# P}$--hard
if some $\mathbf{\# P}$--complete problem is polynomial--time reducible to it. Other instances of 
$\mathbf{\# P}$--complete problems are the counting of Hamiltonian paths in a graph \cite{GaJo} and the more intractable
problems arising in statistical mechanics, such as the enumeration of all configurations contributing to  ground 
state partition functions \cite{Wu}. 
 
The computational intractability of Problem 1 does not rules out  the possibility of `approximating'
efficiently Jones invariant, so that we may ask 
\begin{quote}
{\em \underline{Problem 2}\\
 How hard is it to approximate the Jones polynomial $\mathit{J}(L,q)$ of a link
$L$ at a fixed root of unity $q$ ($q \neq$ 2nd, 3rd, 4th, 6th root)?}
\end{quote}
Loosely speaking, the approximation we are speaking about is a number $Z$ such that, for any choice
of a small $\eta > 0$, the numerical value of $\mathit{J}(L,q)$, when we
substitue in its expression the given value of $q$,  differs from $Z$ by an amount
ranging  between $-\eta$ and $+\eta$ (see section 3.2 below for a more precise statement).
In the framework of classical complexity theory no algorithm to handle 
efficiently Problem 2 exists, while 
the answer in the quantum computational context was given in \cite{BoFrLo} (see also \cite{WoYa}): 
\begin{quote}
 The approximation of the Jones polynomial of a link presented as the closure of a braid
at any fixed root of unity is $\mathbf{BPQ}$--complete. Moreover, this problem is universal
for quantum computation, namely is the `prototype' of all
problems efficiently solvable on a quantum computer.
\end{quote}
Recall that
$\mathbf{BQP}$ is the computational complexity class of 
problems which can be solved in polynomial time by a quantum computer 
with a probability of success at
least $\frac{1}{2}$ for some fixed (bounded) error. In \cite{BoFrLo} it was
proved that $\mathbf{P}^{\mathit{J}}$ $=\mathbf{BQP}$, where $\mathbf{P}^{\mathit{J}}$ is defined as the 
class of languages accepted
in polynomial time by a quantum Turing
machine with an oracle for the language defined by  Problem 2. This equality between 
computational  classes implies that, if we find out  an efficient quantum algorithm
for Problem 2, then the problem itself is complete for the class $\mathbf{BQP}$, 
namely  each problem in this class can be efficiently reduced to 
a proper approximate evaluation of the Jones polynomial of a link \cite{WoYa}.

Explicit, efficient quantum algorithms for approximating the Jones polynomial
were proposed by Aharonov, Jones and Landau \cite{AhJoLa} and by Wocjan and Yard \cite{WoYa},
while an early attempt can be found in \cite{SuRa}.\\
In our papers \cite{GaMaRa1,GaMaRa2} we proved that efficient algorithms can be implemented  for
approximating  the larger class of `colored' Jones polynomials of links
(addressed also in \cite{KaLo}).

The issue of colored link invariants brings us back to the quantum CSW environment,
where they represent the most general gauge invariant physical observables of the
theory, being vacuum expectation values of generic Wilson loop operators \cite{Wit,KaGoRa,Gua}.
In section 3 we shall provide a generalization of our algorithm for colored
polynomials to handle the quantum $3$--manifold invariants introduced in 1.1.

\vspace{12pt}

Summarizing our results, we have shown that all the significant quantities --partition functions and
observables-- in $SU(2)$ quantum CSW theory can be efficiently approximated at finite values of the
coupling constant $k$. The intrinsic field--theoretic solvability of CSW theory is thus reflected by its
computability on a quantum computer. Looking at the question 
the other way around, classical computational intractability
of Jones and colored polynomials can be viewed as a consequence of their quantum nature. This feature 
has prevented up to now both exact and approximate efficient calculations of such topological quantities on classical  
(probabilistic) machines as it happens for simulations of any non trivial `genuine' quantum mechanical system 
(see {\em e.g.} Feynman's proof in \cite{Fey}).

\subsection{Quantized
 geometry {\em versus} quantum computing}

In this section we address some implications of our results
in connection with algorithmic questions that may arise in other physical theories whose
dynamical variables have a geometric character, typically quantum gravity models in $D=3$
and $4$
spacetime dimensions.\\
We have however to face preliminarly a conceptual dilemma, namely whether
\begin{quote} {\bf i)}
an abstract universal model of computation, able to simulate any discrete quantum system
including solvable topological field theories, must exists by its own
\end{quote}
or
\begin{quote} {\bf ii)}
a (suitably chosen) quantum system is by itself a computing machine whose internal evolution can 
reproduce the proper dynamics of classes of physical systems.
\end{quote}
The second alternative is becoming quite popular thanks to Lloyd's model, where a net of computing units generates 
a (superposition of) $4D$ spacetimes \cite{Llo}  (see also \cite{Ziz} where similar ideas were anticipated).

The idea that many (if not all) aspects of our reality may be thought of as `outputs' of some kind of
information processing is both appealing and intriguing. In this connection the role of information theory
and its tools is so enhanced that it becomes a unifying paradigm. Of course the classical
version of hypothesis {\bf i)} is usually taken for granted as far as, on the one hand,
 a (probabilistic) Turing machine
is capable of simulating the evolution of any classical system within a given accuracy,
and, on the other, all concrete, finite--size realizations of the abstract machine obey the laws of
classical physics. The praticability of hypothesis {\bf ii)} depends heavily
on which system is chosen as a simulator and which types of boundary or initial conditions must
be imposed to reproduce the dynamical behavior of observed physical systems. Moreover, the concept of `efficient'
processing of information seems difficult to be handled without an abstract reference model of computation.

With the previous remarks in mind, we favour assumption {\bf i)}, where the reference model of
computation can be represented by the spin network simulator \cite{MaRa1,MaRa2,MaRa3} 
(see also \cite{MaRa4}) or
its $q$--deformed, finite--size counterpart \cite{GaMaRa1,GaMaRa2} (see also section 2 below).
Note however that it may be tempting to assert that the spin network
--thought of as a real net of interacting spin variables-- can play as well the role of
the reference quantum system in statement {\bf ii)}. This is due to the
fact that the recoupling theory of $SU(2)$ angular momenta --representing the algebraic
substratum of the simulator-- is the main ingredient of the `spin network models'
introduced by Ponzano and Regge \cite{PoRe} and Penrose \cite{Pen}. Here classical, discretized
$3D$ euclidean geometry arises from recouplings of quantum angular momenta in the
asymptotic, large angular momentum limit (see \cite{MaRa2}, section 5 for a brief account). Thus,
much in the sense of {\bf ii)}, spin networks may act --under suitable constraints--
 as computing machines able to process
information encoded into quantum spins to output `quantized' $3$--geometries obeying
Einstein equations in the classical limit.

\vspace{12pt}

Thinking back to the issue of algorithms for quantum invariants defined in the
framework of quantum CSW theory, our results can be used to test the algorithmic
complexity of quantum gravity models in $D=3$ spacetime dimensions too. This 
achievement can be justified --independently from spin network models--
by exploiting the close connection between CSW theory and $3D$ gravity both
as classical field theories and at the quantum level \cite{Wit,Car1} 
(classical $3D$ gravity with a positive cosmological constant, reinterpreted as
an $SU(2)$ gauge theory, is quantized through the Euclidean path integral prescription).\\
For a closed orientable Riemannian $3$--manifold ${\cal M}^3$, let
$Z( {\cal M}^3;k)$ denote the Witten partition function associated with the classical
CS action $S_{CS}(A)$, $A$ being the connection $1$--form. The functional
$Z( {\cal M}^3;k) \overline{Z({\cal M}^3;k)}$ $\equiv |Z( {\cal M}^3;k) |^2$ $=
{\cal Z}({\cal M}^3;k)$ for finite $k$ is the partition function of
$3D$ Euclidean quantum gravity in the first--order form, where
the coupling constant $k$ is related to the cosmological constant $\Lambda$ by
$k=4 \pi /\sqrt{\Lambda}$. 
\footnote{Note in passing that the invariant ${\cal Z}({\cal M}^3;k)$,
on the one hand,  equals the Turaev--Viro invariant 
for triangulated $3$--manifolds \cite{TuVi} and, on the other, can be derived
by relating $3D$ gravity to a BF--type topological field theory \cite{Wit,KaGoRa,Car1}. }\\
The point here is that every manifold in the class considered here can be presented
as the the complement of a link $L$ in the $3$--sphere (technically, by surgery
along a framed version of $L$, see section 3.1), 
so that ${\cal M}^3 \equiv {\cal M}^3_L$ $\sim (S^3 \setminus L)$
and ${\cal Z}({\cal M}^3_L;k)$ $=|Z({\cal M}^3_L;k)|^2 $, where  $Z({\cal M}^3_L;k)$
is the Reshetikhin--Turaev--Witten quantum invariant introduced in 1.1.
Thus the results on algorithms for quantum invariants of links and $3$--manifolds
summarized in 1.2 work equally well in the context of Euclidean $3D$
quantum gravity models, reflecting once more the  exact solvability of
the theory for finite $k$.

\vspace{12pt}

An even more interesting connection between $SU(2)_q$ quantum invariants and quantum gravity
emerges when dealing with canonical quantization methods applied to general relativity in
$(3+1)$ dimensions. We refer in particular to the `loop representation' based on Ashtekar
`connection representation' of canonical gravity (see \cite{GaPu,Rov} for reviews
and original references). The Reshetikhin--Turaev--Viro invariants act there
as quantum states associated with the boundary $3$--geometries (spatial slices of
$4D$ spacetimes). Such states were shown to satisfy both the quantum diffeomorphism constraint
and the quantum Hamiltonian constraint (Wheeler--DeWitt equation) once a cut--off --related
to the deformation parameter $q$-- has been fixed. Thus quantum invariants constitute
the natural kinematical arena of loop quantum gravity \cite{GaGrPu}.
This sketchy scenario hides however the necessity of introducing new types of quantum invariants
--the perturbative invariants-- which arise when the cut--off of the underlying
quantum field theory is removed.

\vspace{12pt}

In the CSW field--theoretic setting, perturbative invariants emerge as
coefficients of the asymptotic  expansion of the partition function $Z({\cal M}^3_L;k)$
as $k \rightarrow \infty$ (see \cite{Oht}, ch. 7). Besides a first term that
corresponds to the semiclassical approximation of $Z({\cal M}^3_L;k)$ (saddle point),
each contribution in the expansion is $\exp\{2\pi k S_{CS}(\hat{A})\}\,$ times a power series in $1/k$
(here $S_{CS}(\hat{A})$ is the CS action evaluated for the flat connection $\hat{A}$). 
Perturbative invariants are the coefficients of the powers $(1/k)^n$ evaluated by using $(n+1)$--loop
Feynman diagrams. No complete perturbative treatment of quantum CSW theory is available at present,
and the meaning of such invariants in geometric topology is only conjectured in a few cases.\\
We are interested here in discussing briefly the `volume conjecture' concerning 
special classes of hyperbolic $3$--manifolds (recall that the volume
is a topological invariant for such manifolds). Note also that
most manifolds obtained by surgery on framed links in the
$3$--sphere can be endowed with hyperbolic metrics. 
Focusing in particular on `hyperbolic knots', namely 
those knots which give rise  to finite volume 
hyperbolic $3$--manifolds, the volume conjecture proposed 
in \cite{Kas,MuMu} 
(see also the review \cite{Oht}
for extended versions) can be cast in the form
\begin{equation}\label{volcon}
2\pi\,\lim_{N \rightarrow \infty}\;\frac{\log |\mathit{J}_N (K)|}{N}\;=\;
\mbox{Vol} \,(S^3 \setminus K)\,,
\end{equation}
 where $K$ is a hyperbolic knot and the notation $\mathit{J}_N (K)$
 stands for the $N$--colored  polynomial of $K$ evaluated at $q=\exp (2\pi i/N)$.\\
As pointed out many times, all  quantum algorithms dealing with link polynomials
 are established for a fixed choice of the root of unity $q$ appearing in the argument
 of the invariants, while the volume
conjecture involves the analysis of the asymptotic behavior of single--colored
polynomials of a same knot for increasing values of the coloring itself.

It would be interesting to explore the possibility of
borrowing some of the techniques  employed in \cite{AhAr} 
for dealing with the asymptotics of the Jones polynomial
to test conjecture (\ref{volcon})
within the computational framework for colored polynomials
addressed in the present paper.  

Note finally that the relevance of this issue for quantum gravity stems from the observation
that all vacuum solutions of Einstein field equations in $(2+1)$ dimensions
with a negative cosmological constant are hyperbolic metrics. As recently shown by
Carlip \cite{Car2}, the smallest hyperbolic volumes give rise to the largest contributions in the
saddle point term of the path integral of the quantum theory.
This implies in turn that the so--called `real tunneling geometries' are most probable 
(a real tunneling geometry represents the transition from a compact Riemannian spacetime
to a Lorentian one within the framework of the Hartle--Hawking `no boundary'
approach to quantum cosmology \cite{HaHa}). On the other hand,
the possibility of controlling such invariants from the algorithmic point of
view might help also in selecting weights to be assigned to $3$--geometries in the
`sum over topologies' within a `fully quantum' cosmology theory.


\section{Processing braiding operators}


\subsection{$q$--deformed spin network and quantum recognizers}

As pointed out in the introduction, our reference 
model of computation to deal with quantum topological invariants
derived from Kaul unitary representations of colored oriented braids \cite{RaGoKa,Kau1}
is given by the $q$--deformed spin netwok model.\\
The (undeformed) spin network simulator has been extensively 
addressed elsewhere \cite{MaRa2} but for the convenience of the reader 
we have collected in appendix A a few mathematical details. 
However, in order to
recognize the necessity of introducing a $q$--deformed 
version, it is worth to discuss here the (categorical) foundations
of the quantum theory of angular momenta.
The computational space of the spin network simulator
--modelled as a graph-- encodes the representation ring
of the Lie group $SU(2)$ --namely finite--dimensional Hilbert spaces
supporting irreducible representations (irreps) of $SU(2)$ endowed 
with  two binary operations, tensor product  $\otimes$ 
and direct sum $\oplus$ (that provide a ring 
structure over the field $\mathbb{C}$)-- together
with all the unitary operators relating (multiple tensor products of) 
such spaces. 
Unlike the usual quantum
circuit model \cite{NiCh}, here it is possible to handle directly eigenstates
of $N$  binary coupled angular momentum variables labelled by
integers and half--integers $j_1,j_2,\ldots,j_N$ (in $\hbar$ units) and
not simply $N$--qubit states labelled by the fundamental irrep
 $j_1=j_2= \ldots =j_N=1/2$. The (re)coupling theory of $N$ $SU(2)$
 angular momenta provides the whole class of unitary transformations that can be 
 performed on many--body quantum systems described by pure angular
 momentum binary coupled sets of eigenstates (\cite{BiLo9}, topic 12). 
The unitaries that we need  here
 (referred to as $j$--gates in section 3 of \cite{MaRa2}) are phase transformations,
 related to swaps of two contiguous spins, and
 recoupling trasformations expressed in terms of $3nj$--coefficients of $SU(2)$
and related to  changes in the interaction schemes of the 
$N$ angular momenta ($N=n+1$). 

Within the framework of the categorical approach,
the $SU(2)$ representation ring is an instance of a unitary tensor category,
endowed with intertwiner spaces and two basic morphisms, a `twist' (a trivial type of `braiding') 
and an `associator' \cite{FuSc}, \cite{Joy}. The former acts on the tensor product
of two Hilbert spaces $V, W$ supporting irreps of $SU(2)$ by exchanging 
the order of the factors, namely
\begin{equation}\label{twist}
R_{V,W}\,:\;V\,\otimes\,W\;\rightarrow\;W\,\otimes\,V
\end{equation}
$$ \text{with}\;\;\;R_{W,V} \circ R_{V, W}\,=\text{Id}_{V \otimes W}.
$$ 
The explicit action of $R$
  on a (binary coupled) state is a trivial phase transform, see \eqref{8},
\eqref{9} in appendix A.\\
  The associator $F$ relates different binary bracketing structures
  in the triple tensor product of irreps $V,U,W$
 \begin{equation}\label{assoc}
F\,:\;(V\,\otimes\,U)\,\otimes\,W\;\rightarrow\;V\,\otimes\,(U\,\otimes\,W)
\end{equation}
and is implemented on a binary coupled state as a Racah transform involving one Racah--Wigner
$6j$--symbol (see \eqref{6}, \eqref{7} in appendix A). Note that both \eqref{twist} and
\eqref{assoc} are isomorphisms (unitary morphisms between intertwiner
spaces in the categorical language) but the associator reflects a true (physically
measurable) modification of the way in which intertwiner spaces are coupled.
 
The remarkable fact, derived from the general theory
of braided tensor categories \cite{JoSt},  is that
more complicated multiple tensor products can be handled without introducing
any further independent morphism. Actually, multiple
tensor product spaces can be related by different combinations of 
braidings and associators, so that the basic morphisms must satisfy
compatibility conditions, a so--called pentagon identity
and two exagon identies. 
In the more concrete language of $SU(2)$ recoupling theory,
each $3nj$--symbol can be obtained as a combination of 
phase and Racah transforms and the non--uniqueness of 
such decomposition is translated 
into the Biedenharn--Elliott (pentagon) identity and the Racah
identity (see {\em e.g.} \cite{BiLo9} or \cite{Russi}
for their explicit expressions).

\vspace{12pt}

In order to deal
with non--trivial braiding operators --to be used in connection
with the study of braid group representations and braid statistics 
(typically occuring in anyonic systems)--
we are forced  to modify
the operation $R$ in \eqref{twist}
by introducing a genuine, non--trivial braiding   morphism \cite{JoSt}
\begin{equation}\label{braidmor}
\mathcal{R}_{V,W}\,:\;V\,\otimes\,W\;\rightarrow\;W\,\otimes\,V
\end{equation}
$$ \text{with}\;\;\;\; \mathcal{R}_{W,V} \circ \mathcal{R}_{V, W} \neq
\text{Id}_{V \otimes W}. 
$$
A consistent way of modifying the $SU(2)$ tensor category 
to include non--trivial braidings can be achieved  
by  moving to the representation ring $\mathfrak{R}$ $(SU(2)_q\,)$
of the $q$--deformed 
Hopf algebra of the Lie group $SU(2)$, $SU(2)_q$ ($q$ a root of unity).
The resulting braided tensor category is  the `universal' algebraic
structure underlying the constructions of quantum
invariants of links and $3$--manifolds outlined in 1.1 and 1.2.
(We refer the reader to \cite{GaMaRa1} and to 
\cite{GaMaRa3}  for short 
technical surveys of the quantum group and CSW approaches.).

According to the above remarks, it should be clear that
the `$q$--deformed' spin network model of 
computation \cite{GaMaRa1,GaMaRa2} is modelled on the $q$--tensor category
\begin{equation}\label{qcategory}
(\,\mathfrak{R} (SU(2)_q\,)\,;\,\mathcal{R}\,;\,\mathcal{F}\,),
\end{equation}
where we have denoted by $\mathcal{F}$ the $q$--counterpart
of the associator $F$ in \eqref{assoc}. Once suitable basis sets are
chosen in the finite collection of irreducible spaces $\in\, \mathfrak{R} (SU(2)_q\,)$,
the unitary morphisms $\mathcal{R}$ and $\mathcal{F}$ can be made
explicit. In particular $\mathcal{F}$ turns out to contain the 
$q$--deformed counterpart of the $6j$--symbol and, regarding it 
as a unitary matrix, it is also referred to as `duality' (or
`fusion') matrix borrowing the language of conformal field theories \cite{GoRuSi}.

\vspace{12pt}

The efficient quantum algorithm for
(approximating) $SU(2)_{q\,}$--colored link polynomials
we obtained in \cite{GaMaRa2}  relies on a two--level
procedure which can be summarized as follows.
\begin{itemize}
\item Kaul unitary representation of colored oriented braids
--associated with links  presented as plat closures
of such braids-- is processed on the $q$--spin network 
\eqref{qcategory} in a number of steps that grows polynomially
in the size of the input. In particular, each elementary
computational step is implemented by applying either $\mathcal{R}$
or $\mathcal{F}$ (see section 2.2 below).
\item The basic $q$--morphisms $\mathcal{R}$, $\mathcal {F}$
in the Kaul representation can be efficiently
compiled on a standard quantum computer, by means of universal elementary
gates acting on suitable qubit--registers. 
\footnote{This result is quite interesting by itself as recently pointed out in  
\cite{FrWa} where relations between the basic morphisms and 
large Fourier transforms are addressed.\\
The problem of finding out efficient algorithms to compute the fundamental 
functions of the quantum theory of $SU(2)$ angular momenta
--Clebsch--Gordan coefficients, $6j$--symbols-- has attracted much attention
\cite{BaHaCh}. However, to our knowledge, there are neither
classical nor quantum algorithms avalaible for evaluating these functions for 
arbitrary values of their arguments. The crucial remark is
that a $q$--$6j$--symbol with arbitrary entries 
can be efficiently compiled and approximated on a
quantum circuit independently of the input size 
of the algorithmic problem owing to the presence
of the natural cut--off provided by $k$ (see section 3.2,
in particular footnote 8).
Then we realize once more that
the $q$--symmetry of solvable topological field
theories is indissolubly tied with the effective computability
of these models.}
\end{itemize}
In order to analyze in more details the first topic above
we need to frame 
the $q$--deformed spin network model
within the theory of
quantum automata and quantum languages.
According to \cite{WiCr} a 
{\em quantum recognizer} 
is a particular type of finite--states quantum machine
defined as a 5--tuple
$\{Q, \mathcal{H}, X, Y,$ ${\bf T}(Y|X)\}$, where
\begin{enumerate}
\item $Q$ is a set of ${\mathfrak n}$ basis states, the {\em internal states};
\item ${\mathcal H}$ is an ${\mathfrak n}$--dimensional Hilbert space
and we shall denote by $|\Psi_0\rangle \, \in \mathcal{H}$ a start state expressed in the given basis;
\item $X$ and $Y= \{ \text {accept, reject}, \epsilon \}$ are finite alphabets 
for input and output symbols respectively ($\epsilon$ denotes the null symbol);
\item ${\bf T}(Y|X)$ is the subset of ${\mathfrak n} \times {\mathfrak n}$ transition matrices
of the form 
$\{\,T(y|x)\,=\,U(x)P(y);\; x \in X,\, y \in Y\}$, 
where $U(x)$ is a unitary matrix
which determines the  state vector evolution and $P(y)$ is a projection operator 
associated with the output measurement on (suitable complete
sets of observables  associated with)  the upgraded state vector.
\end{enumerate}
In this kind of machine the output alphabet is chosen in such a way that
a word $w$ written in the input alphabeth $X$ must be either accepted or rejected, 
while for the null symbol the requirement is
$P(\epsilon)\equiv {\mathbb I}$ (the identity matrix).  
Thus the one--step transition matrices applied to the start state $|\Psi_0 \rangle$ 
can in principle assume the forms
\begin{itemize}
\item[a)] $T(\epsilon|x)=U(x)P(\epsilon)\equiv U(x){\mathbb I} \qquad \forall x \in X$,
\item[b)] $T(\text{accept}|x)=U(x)P(\text{accept})\qquad \forall x \in X$ with $P(\text{accept})
\equiv |\Psi_0\rangle \langle \Psi_0|$,
\item[c)] $T(\text{reject}|x)=U(x)P(\text{reject})\qquad \forall x\in X$ 
with $P(\text{reject})\equiv {\mathbb I}-
|\Psi_0\rangle \langle \Psi_0|$,
\end{itemize}
according to whether no measure is performed (case a)), or the output 
is `accept'/`reject', namely cases b) /c) respectively.

The general axioms stated above can be
suitably adapted to make this machine able to recognize  
a language $\mathcal{L}$ 
endowed with a word--probability 
distribution $\mathfrak{p}(w)$ over the set of words 
$\{w \} \in \mathcal{L}$.
In particular, 
for any word $w=x_1x_2\dots x_l \in \mathcal{L}$ 
the recognizer one--step transition matrix elements 
are required  to be of the form 
$T_{ij}(x_s)=U_{ij}(x_s)$ 
on reading each individual symbol $ x_s \in w$,
namely no measurement is performed at the intermediate steps
(here $i,j$ run from 1 to
$\mathfrak{n}$, the dimension of the Hilbert space $\mathcal{H}$).
Each $U_{ij}(x_s)$ must
satisfy the condition 
\begin{equation}
\label{poscond}
|\,U_{ij}(x_s)\,|^2\,>\,0\,,
\end{equation}
and the recognizer upgrades the (normalized) initial state to
\begin{equation}
\label{recseq}
U(w)\,|\,\Psi_0\rangle \,\doteq\, U(x_l)\dots U(x_1)\,|\,\Psi_0\rangle\,.
\end{equation}
Then the machine assigns to the word $w$ the number
\begin{equation}\label{autpro}
\mathfrak{p}(w) \,=\,|\,\langle \Psi_0|\,U(w)\,P(\text{accept})\,U(w)|\Psi_0\rangle\,| 
\;\;\text{with}\;\;\; 0 \leq \mathfrak{p}(w) \leq 1\,,
\end{equation}
which corresponds to the probability of accepting the 
word $w$ as a whole.

More generally, the machine accepts a word $w$
according to an {\em a priori} probability distribution ${\text Pr}(w)$
with
a word--probability treshold $0 \leq \delta \leq 1$ 
if
\begin{equation}\label{tresh}
|\,{\text Pr}(w)\,-\,\mathfrak{p}(w)\,|\; \leq \,\delta\,,\;\;\;\forall w \in \mathcal{L}\,.
\end{equation}
In what follows the accuracy $\delta$ will be set to $0$, so that
the two probability distributions ${\text Pr}$ and $\mathfrak{p}$ coincide.


\subsection{The Kaul representation as a quantum language}
 
In this section we shall show that 
the $q$--deformed spin network computational
scheme  \eqref{qcategory} embodies
families of quantum finite--states machines 
(or quantum automata)
$\{\mathcal{A}_q\}$ 
-- parametrized by the labels of $N$ irreps of
$SU(2)_{q\,}$, $q=$root of unity--
that recognize the language generated by the 
braid group according to a probability 
distribution given by the square of the modulus of
the $q$--colored link polynomial. \\
This construction complies  essentially
with what we have done in \cite{GaMaRa1}, but 
here we should stress the interpretation
of the Kaul representation \cite{Kau1}
as a quantum language, on the one hand,
and  the role played by the probability distribution, on the other
(leaving aside details on the field--theoretic background material).

\begin{figure}[htbp]
	\centering
		\includegraphics[width=0.10\textwidth]{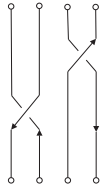}
		\caption{An oriented braid on four strands.}
	\label{fig:orientedb}
\end{figure}

The basic ingredients of Kaul's construction are oriented geometric braids
(see Fig. [\ref{fig:orientedb}]), the strands of which are endowed 
with `colorings' given by $SU(2)_q$ irreps labels.
An $n$--strand colored oriented braid is defined by two sets of
$n$ assignments $\hat{j}_i=\left( j_i,\varepsilon_i \right)$ with 
$(i=1,2,\dots,n)$, corresponding to the spin $j_i$ labelling the 
strand and to the orientation $\varepsilon_i$ of the strand, with 
$\varepsilon_i=\pm 1$ (for the 
strand going into or away, respectively, 
from a horizontal rod from which the braid issues). The first set of 
assignments is associated to the upper rod, the second  to 
the lower rod (we use the convention that 
two braids are composed in the downward direction). 
The conjugate of $\hat{j}_i$ is defined as 
$\hat{j}^*_i \equiv (j_i, -\varepsilon_i)$. It follows 
that the assignments on the lower rod are just 
a permutation of the conjugates of the assignments on the 
upper rod. A colored and oriented braid can thus be 
represented by the symbol 
\begin{equation}\label{colorbr}
\sigma \left( 
\begin{array}{cccc}
\hat{j}_1 & \hat{j}_2 & \dots & \hat{j}_n\\ 
\hat{l}_1 & \hat{l}_2 & \dots & \hat{l}_n
\end{array}
 \right),
\end{equation}
where $\hat{l}_j=\hat{j}_{\pi(i)}$ for some $i$ and $j$ and  a permutation $\pi$
of $\{1,2,\ldots n\}$.

\begin{figure}[htbp]
	\centering
		\includegraphics[width=0.15\textwidth]{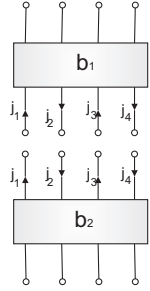}
		\caption{The composition of two colored oriented braids.}
	\label{fig:composition2}
\end{figure}

\begin{figure}[htbp]
	\centering
		\includegraphics[width=0.40\textwidth]{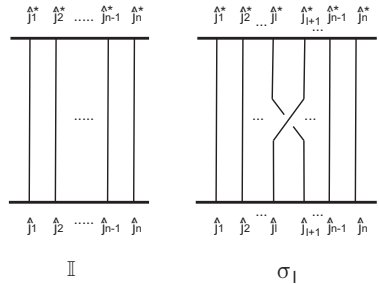}
		\caption{The identity and the generator $\sigma_l$
                 of the colored oriented braid group.}
	\label{fig:colgenerators}
\end{figure}

The composition of two colored oriented braids is well 
defined only if the orientations and the colors of the 
two braids match at the merging points, as shown in 
Fig. [\ref{fig:composition2}].
The group of colored oriented braid is generated by the identities 
(one for each assignment of colors and orientations on a 
topologically unentangled braid) and by the braids of type 
$\sigma_l$, as shown in Fig. [\ref{fig:colgenerators}].

\begin{figure}[htbp]
	\centering
		\includegraphics[width=0.50\textwidth]{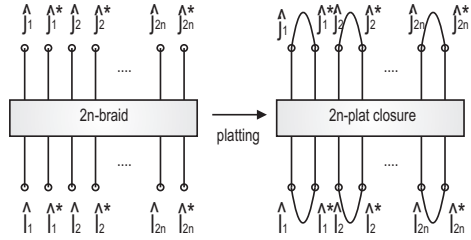}
		\caption{The platting of a colored oriented braid on $2n$ strands.}
	\label{fig:platting}
\end{figure}

The collection of $\{\sigma_l\}$ for $l=1,2,\ldots n-1$
are the standard generators of the (colored) braid group $B_{n}$ and
satisfy  the following defining relations
\begin{eqnarray}\label{brgrgen}
\sigma_i \sigma_{i+1} \sigma_i &=& \sigma_{i+1} \sigma_{i} \sigma_{i+1}\nonumber \\ 
\sigma_i \sigma_j &=& \sigma_j \sigma_i \qquad |i-j|\geq 2.
\end{eqnarray}
The inverse of a generator $\sigma_l$, $(\sigma_l)^{-1}$, 
corresponds
to the under--crossing of the
left strand $\hat{j}_l$ in Fig. [\ref{fig:colgenerators}] (right).\\ 
In order to obtain a link (multicomponent knot) from a colored braid we 
need to `close up' the  braid. For our purposes we may consider only the plat closure 
(or {\em platting}) of a colored braid, 
defined for braids which possess  an even number of 
strands and whose assignments match as follows (see also Fig. [\ref{fig:platting}] )
\begin{equation}\label{orcolbr}
\sigma\left( 
\begin{array}{ccccccc}
\hat{j}_1&\hat{j}_1^*&\hat{j}_2&\hat{j}_2^*&\dots & \hat{j}_{2n}&\hat{j}_{2n}^*\\
\hat{l}_1&\hat{l}_1^*&\hat{l}_2&\hat{l}_2^*&\dots & \hat{l}_{2n}&\hat{l}_{2n}^*
\end{array}
 \right).
\end{equation}

Since any (colored oriented) link can be obtained  
as the plat closure of a  braid \cite{Bir},   
we do not lose generality for what concerns the class of
links that can be handled.

\begin{figure}[htbp]
	\centering
		\includegraphics[width=0.50\textwidth]{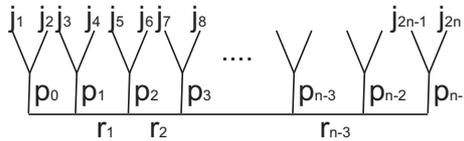}
		\caption{The conformal block of type $\{ {\bf p};{\bf r} \}$ (odd).}
	\label{fig:confodd}
\end{figure}

\begin{figure}[htbp]
	\centering
		\includegraphics[width=0.50\textwidth]{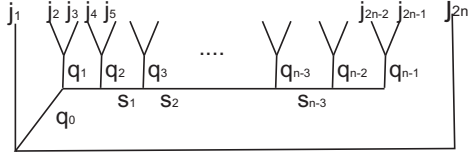}
		\caption{The conformal block of type $\{ {\bf q};{\bf s} \}$ (even).}
	\label{fig:confeven}
\end{figure}

The further step consists in embedding the $2n$--strand braid into
a $3$--sphere $S^3$ with two three--balls removed, giving rise
to a $3$--manifold with two boundaries $\Sigma^1, \Sigma^2$ (topologically
two  $2$-spheres $S^2$ with opposite orientations). 
The intersections (`punctures')  of the braid \eqref{orcolbr}
with the boundaries inherit the colorings and orientations from
the corresponding strands of the braid (to be associated with
Wilson line operators in the ambient CSW topological field theory).
Following \cite{Wit}, finite dimensional 
Hilbert spaces ${\mathcal H}^1 \otimes {\mathcal H}^2$ are associated with
to the two boundaries $\Sigma^1, \Sigma^2$, and the basis
sets in these spaces are the so--called conformal blocks
of the boundary Wess--Zumino--Witten conformal field theory at level $\ell$, 
with $2n$ external lines labelled by (different) irreps of $SU(2)_q$ 
($\ell$ is related to CSW coupling constant by $\ell =k+2$,
so that from now on we set $q= \exp \{2\pi i/\ell\}$).
Two particular types  of 
conformal block bases are needed to deal with braids the plat closures of
which will give rise to colored oriented links, and
their combinatorial patterns are shown in Fig. 
[\ref{fig:confodd}] and Fig. [\ref{fig:confeven}].

The (orthonormal) basis sets are constructed by taking particular 
binary coupling schemes of the $2n$ `incoming' angular momentum variables
$j_1,j_2,\ldots, j_{2n}$ which must sum up to give a  spin--$0$ total 
singlet state.\footnote{Incidentally, such a choice complies also with the
conditions necessary for the construction of error avoiding codes and implies
as well  the robustness of the scheme \cite{ZaRa,PaZaRa}.}
 The procedure can be carried out by parallelling the $SU(2)$, 
undeformed case (see appendix A) but
here the labels of the irreps $j_i$'s (integer and half--integers) are constrained to range
from $0$ to $\ell/2$ and the binary bracketings on tensor products are decomposed
according to rules of $SU(2)_{q\,}$ representation theory
(see {\em e.g.} \cite{GaMaRa3}, section 3).

Looking at  the combinatorial structure
of Fig. [\ref{fig:confodd}], the most general odd--coupled basis is
consistently labelled as
\begin{equation}\label{oddbas}
|\,\mathbf{p}; \mathbf{r}\,\rangle^{\mathbf{j}}\;\,,
\end{equation}
where $\mathbf{j}$ stands for the ordered 
string $j_1,j_2,\ldots j_{2n}$, $\mathbf{p} \equiv p_0, p_1, \dots, p_{n-1}$
and $\mathbf{r} \equiv r_1, r_2,\dots, r_{n-3}$.
In the even--coupled case depicted in Fig. [\ref{fig:confeven}],  
the states of the basis are denoted by 
\begin{equation}\label{evebas}
|\,\mathbf{q}; \mathbf{s}\,\rangle^{\mathbf{j}}\;\,,
\end{equation}
where $\mathbf{j}$ is the same as before while 
$\mathbf{q} \equiv q_0, q_1, \dots, q_{n-1}$
and $\mathbf{s} \equiv s_1, s_2,\dots, s_{n-3}$. 

\begin{figure}[htbp]
	\centering
		\includegraphics[width=0.30\textwidth]{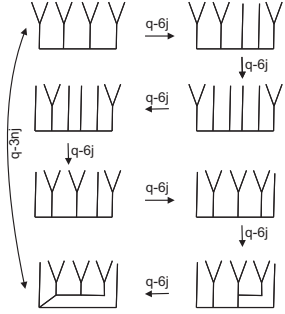}
		\caption{Example of decomposition of a duality transformation
 between the two  extremal conformal blocks in the case of eight incoming spins.}
	\label{fig:decomposition}
\end{figure}

\begin{figure}[htbp]
	\centering
		\includegraphics[width=0.30\textwidth]{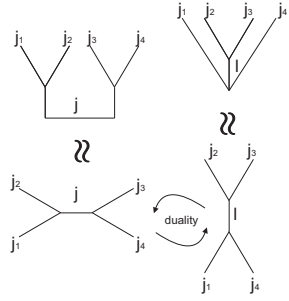}
		\caption{The elementary duality transformation associated with a $q$--$6j$ symbol.}
	\label{fig:duality2}
\end{figure}

The basis vectors associated to the conformal blocks 
 \eqref{oddbas} and \eqref{evebas} are related to each other by 
\begin{equation}\label{3nj}
|{\bf p};{\bf r} \rangle ^{{\bf j}}=
\sum_{({\bf q};{\bf s})} 
A_{({\bf p};{\bf r})}^{({\bf q};{\bf s})} 
\left[ 
\begin{array}{cc}
j_1&j_2\\ 
j_3&j_4\\ 
\vdots & \vdots \\ 
j_{2n-1}&j_{2n}
\end{array}
 \right] 
 | {\bf q};{\bf s} \rangle^{{\bf j}},
\end{equation}
where the symbol $A_{({\bf p};{\bf r})}^{({\bf q};{\bf s})}\,[: \, : ]$ 
represents the unitary duality matrix (or $q$--deformed $3nj$ recoupling coefficient).
As pointed out in section 2.1, it is
a standard result that any such duality matrix can be decomposed into
(sums of) products of 'basic' duality matrices or $q$--$6j$ symbols,
see section 3.2 below.\\
A  graphical representation of the decomposition \eqref{3nj} in the case
of eight incoming spin variables is shown  in Fig.[\ref{fig:decomposition}].
Note that the graphical representation of the basic duality transformation 
(the matrix counterpart of the associator
in the language of tensor categories) can be also drawn in the
most familiar form shown in Fig. [\ref{fig:duality2}].

As proved in \cite{Kau1}, the colored polynomial of a link $L$, presented as 
the plat closure $\hat{\sigma}$ of a colored braid  
$\sigma \left( 
\begin{array}{ccccc}
\hat{j}_1 & \hat{j}_1^*& \dots & \hat{j}_{n} & \hat{j}_{n}^*\\ 
\hat{l}_1 & \hat{l}_1^*& \dots & \hat{l}_{n} & \hat{l}_{n}^* 
\end{array}
 \right)$ as defined in \eqref{orcolbr}, 
 is given by 
 \begin{equation}\label{colJon}
 \mathit{J}\,[L;\,\mathbf{j};\, q]\,=\,\text{Tr}\,\prod_{i=1}^{n}
\left[ 2j_i+1 \right]_q \;
 ^{{\bf l}} \! 
 \langle {\bf 0};{\bf 0} |\, 
\hat{U} \left[ \sigma \left( 
\begin{array}{ccccc}
\hat{j}_1 & \hat{j}_1^*& \dots & \hat{j}_{n} & \hat{j}_{n}^*\\ 
\hat{l}_1 & \hat{l}_1^*& \dots & \hat{l}_{n} & \hat{l}_{n}^* 
\end{array}
 \right) \right]
 | {\bf 0};{\bf 0} \rangle^{{\bf j}}\,,
 \end{equation}
where $\mathbf{j}\equiv(j_1,j_2,\dots,j_{2n})$, $[2j_i+1]_q$ is  
the quantum dimension of the irrep labelled by $j_i$ and 
the quantum integer $[x]_q$ is defined as
\begin{equation}\label{qdim} 
[x]_q \equiv \frac{x^{q/2}-x^{-q/2}}{x^{1/2}-x^{-1/2}}.
\end{equation}
Thus $\mathit{J}\,[L;\,\mathbf{j};\, q]$ can be evaluated by taking the  trace of the  matrix elements
of the composite braiding operator $\hat{U}[\sigma]$ in the Kaul representation
with respect to the odd--coupled basis, where all the
intermediate quantum numbers are  constrained to give singlet  
eigenstates (a similar result would hold true for the even--coupled 
basis).Moreover, $\hat{U}[\sigma]$ can be decomposed into a finite sequence of unitary matrices
$U[\sigma_{2l+1}]$ (diagonal matrices in the odd--coupled basis
adopted in \eqref{colJon}) and duality matrices of the type
\eqref{3nj} to be applied whenever a switch to the even--coupled
basis is needed, namely when an even $U[\sigma_{2l}]$ occurs in the decomposition
(see \cite{Kau1} for the explicit expression of these matrices). 

\vspace{10pt}

The construction outlined above can be cast into an effective
process of calculation  by resorting the concept of quantum recognizer
introduced in section 2.1.\\
The $\mathcal{A}_{q\,}$ recognizer is defined,
for a fixed root of unity $q= \exp \{ 2\pi i/\ell\}$,
 by the 5--tuple 
$\{ \mathcal{C}_{odd},\mathcal{H}, B_{2n},$
$\{ \text{accept, reject}, \epsilon \},$ $U(B_{2n}) \}$, 
where 
\begin{itemize}
\item ${\mathcal C}_{odd}$ is the odd--coupled
conformal block basis 
of the boundary WZW theory (see \eqref{oddbas}
and Fig. [\ref{fig:confodd}]). 
\item ${\mathcal H}$ is the ordered tensor product of $2n$ 
$(2j_i +1)$--dimensional Hilbert spaces supporting irreps 
of $SU(2)_q$ labelled by $j_i$, with $j_i \leq (\ell -2)/2$                
$(i=1,2,\ldots, 2n)$.
\item $B_{2n}$ is the braid group on $2n$ strands whose generators
$g\equiv \{\sigma_1, \sigma_2, \ldots, \sigma_{2n-1}\}$ and their inverses  
represent the input alphabet (we may add the identity
element $e \in B_{2n}$ as null symbol).
\item $Y=\,$ \{ accept, reject, $\epsilon$ \} is the output alphabet.
\item The transition matrices are expressed in terms of
$U(B_{2n})$, denoting collectively the Kaul unitary representation matrices, 
while the projectors $\,P(y)\; (y \,\in \, Y)$
are defined as in the general case given in section 2.1.
\end{itemize}

\noindent According to the above definitions, we provide the automaton
with an input word $w \in B_{2n}$ of length $\kappa$ (written in the alphabeth $g$ by natural
composition in $B_{2n}$)
\begin{equation}\label{brword}
w\,=\,\sigma_{\alpha_1}^{\epsilon_1}\,\sigma_{\alpha_2}^{\epsilon_2} \dotsb 
\sigma_{\alpha_{\kappa}}^{\epsilon_{\kappa}}\,;\;\;\;\;
\;\sigma_{\alpha_i} \in g\,,\epsilon_i=\pm1
\end{equation}
and such that
the (plat) closure $\hat{w}$ of the  $2n$--strand braid  $w$
gives the link $L$ to be processed. Dropping for simplicity
all the matrix indices, the unitary evolution of the 
automaton is achieved by applying the sequence 
\begin{equation}\label{brevol}
U(w)\,=\,U(\sigma_{\alpha_{\kappa}}^{\epsilon_{\kappa}})\,
U(\sigma_{\alpha_{\kappa -1}}^{\epsilon_{\kappa -1}})\dotsb 
U(\sigma_{\alpha_1}^{\epsilon_1})
\end{equation}
to a start ket $| {\bf 0};{\bf 0} \rangle^{\,{\bf j}}$
in the odd--coupled basis \eqref{oddbas}.\\
Whenever an odd--braiding $\sigma_{\alpha}= \sigma_{2i+1}$ (or
$(\sigma_{2i+1})^{-1}$) occurs, the
automaton one--step evolution upgrades the internal state 
by inserting the eigenvalue of the associated unitary $U$.
On the other hand, when an even--braiding $\sigma_{\beta}= \sigma_{2i}$
(or $(\sigma_{2i})^{-1}$)
must be implemented, the automaton has to change  the parity
of the internal state by means of a duality matrix 
\eqref{3nj}, so that the effective transformation is given by
the product $U(\sigma_{\beta})\,A [:\,:]$. Since any duality transformation
can be split into a sequence of basic duality matrices, we may look at 
the $q$--$6j$ symbol as representing  an `elementary', one--step evolution of
the automaton.  By resorting to standard results
in graph theory  it is possible to estimate
how many $q$--$6j$ symbols are needed to decompose the most
general $q$--recoupling coefficient \cite{Belgi}, \cite{GaMaRa1}. In the present case
the upper bound can be expressed in terms of $2n$, the braid index
(or, equivalently, the number of strands of the input braid).
On the basis of the above remarks, the time complexity
function (number of computational steps) for 
processing  a braid--word of length $\kappa$  
on the quantum recognizer $\mathcal{A}_q$ 
is bounded from above by \cite{GaMaRa1}
\begin{equation}\label{stima}
\kappa \; (\tilde{N}\,\ln \tilde{N})\,\;\;\;\;\text{where}\;\tilde{N}\equiv (2n-1),
\end{equation}
implying that the automaton processes efficiently such braids.

Let us finally comment on the `probability distribution'
entering into the definition of a quantum automaton
that recognizes a language in a probabilistic sense
(end of section 2.1). On the basis of the expression of
the colored link invariant given in \eqref{colJon}
and by comparison with the word probability 
of a quantum recognizer defined in \eqref{autpro},
it should be quite clear that the probability
naturally associated with a link $L$ processed on
$\mathcal{A}_q$ is the square modulus
of its colored polynomial
(note that the positivity conditions required in \eqref{poscond}
are always satisfied).

 In order to check this 
result in a concrete case, 
the explicit construction of the $q$--spin network automaton
that recognizes the braid group language with
a probability distribution given by the square 
modulus of the Jones polynomial 
is carried out in appendix B.


\section{Efficient quantum algorithms for $3$--manifold\\ 
quantum invariants}

\subsection{Colored framed links and 3--manifold quantum invariants}

The quantum invariants of $3$--manifolds that we are going to discuss 
--within the mathematical framework developed in \cite{KiMe}, see also \cite{Lic1}--
can be obtained as   combinations of polynomial
invariants of `framed' unoriented links in the $3$--sphere
$S^3$ on the basis of theorem 1 stated below.
It is worth noting that in the CSW environment the necessity
of introducing framings is physically motivated by
the requirement of general covariance of the quantized
field theory (see {\em e.g.} ch. 3 of \cite{Gua}).
 
Loosely speaking, a framed oriented link $[L;\mathbf{f}]$ is obtained from
a link $L$--thought of as made of knotted strings-- by thickening
its strings to get oriented `ribbons'. If $L$  has $S$  
knot components $K_1,K_2,\dots,K_S$, for each 
$K_s$ we introduce another closed path $K_{s}^f$ oriented 
in the same way as $K_s$ and lying
within an infinitesimal neighborhood of $K_s$ (knots and links
are embedded in $\mathbb{R}^3$ or in $S^3$). The overall topology
of the link is not modified, but for each $K_s$ we now have an extra
variable $\tau (K_s)$ telling us how many times the oriented ribbon is `twisted'.
Denoting by $\mathbf{f}\,\doteq\,\{f_s=n(s),\, n(s) \in \mathbb{Z}\}$ $(s=1,2,\dots,S)$ 
the framing of the link $L$, $f_s$ is the self--linking number
of the band, or equivalently the
linking number $lk(K_s,K_s^f)\equiv \chi(K_s,K_s^f)$ between 
the knot $K_s$ and its framing curve $K_s^f$ which winds $n(s)$ times in the right--handed 
direction. 
The twist of the band $\tau (K_s)$ is not independent from
$lk(K_s,K_s^f)$, and the simplest choice we can made  is to set
$$ \tau (K_s)\,=\,w(K_s),
$$
where $w(K_s)$ is the writhe of the $s$--th component.
\footnote{Given a link diagram $D(L)$, namely a projection of the oriented link $L$
onto a plane, we can define two numerical invariants associated with such diagrams. 
The {\em writhe} number $w(D_L)$ is given by
$w(D_L)=\sum_{p} \epsilon(p)$
where $\{p\}$ are the crossing points $\in D(L)$ and  $\epsilon(p)=\pm 1$
according to whether there appears an over--crossing of the left strand over the right strand
or  an under--crossing (both strands are oriented upward).
The {\em linking number}, defined for a link with more than one component knot,
is defined for each pair of components $(K_i,K_j)$ as  $lk(K_i,K_j)$ $=w(D_L)-
w(D_{K_i})- w(D_{K_j})$. \\
It can be shown that the writhe is a regular isotopy invariant
for knots and link and the linking number is an ambient isotopy invariant for links.
Recall that two links in $\mathbb{R}^3$ (or $S^3$) are ambient isotopic if they can be
continuously deformed one into the other. It can be shown that
two links are ambient isotopic if and only if their diagrams
are connected by a finite sequence of Reidemeister moves of
type I, II, II. Regular isotopy is a restricted
type of equivalence among links where the allowed
Reidemeister moves are of type II and III. It is worth noting that 
the colored link polynomials \eqref{colJon} are invariants of
regular isotopy while an associated ambient isotopy invariant can
be obtained by multiplying $\mathit{J}(L;\,\mathbf{j};\,q)$ by
$\{q^{-3w(L)/4}\,/(q^{1/2}-q^{-1/2})\}$, where $w(L)$ is the writhe
defined above.     
}

\begin{figure}[htbp]
	\centering
		\includegraphics[height=5cm]{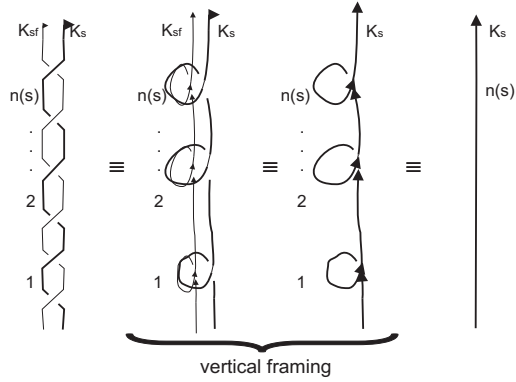}
	\caption{\small{Graphical representations for the vertical framing.}}
	\label{fig:verticalf}
\end{figure}

The type of framing usually adopted is the `vertical' framing, where the 
frame is thought to be placed vertically above the  link diagram. 
Looking at  Fig. [\ref{fig:verticalf}],
the framing can be represented by putting the writhe number $w(K_s)\equiv n(s)$ 
nearby the band 
or even by writing $n(s)$ next to the string representing (a portion of) $K_s$. 

\vspace{10pt}

The $3$--manifolds we are going to consider are closed, connected and oriented,
namely compact and without boundary. In general any such manifold
$\mathcal{M}^3$ can be presented as the union of several components
--endowed with $2$--dimensional boundaries--
glued together by suitable identification prescriptions on the
points lying in their boundaries. If the components are sewed together in
a different way a topologically different manifold $\widetilde{\mathcal{M}^3}$ 
may be obtained (recall that two manifolds are topologically equivalent
iff they are homeomorphic, namely there exists a one--to--one
continuous map between them). The new
manifold $\widetilde{\mathcal{M}^3}$ 
is said to be obtained from $\mathcal{M}^3$ by means of `surgery'.\\
The fundamental theorem that
characterizes (equivalence classes of homeomorphic) $3$--manifolds 
reads \cite{Lic2,Wal}
\begin{quote}
{\bf Theorem 1.}
Every closed, connected and orientable $3$--manifold $\mathcal{M}^3$
can be obtained by surgery on an unoriented framed link in the $3$--sphere
$S^3$.
\end{quote}
Following \cite{Gua}, let us illustrate in some details 
the (Dehn) surgery procedure referred to in the statement of the theorem
in the simple case of a knot $K$ ($1$--component link) $\subset S^3$ .
The building blocks of the construction are solid tori, so let us 
begin the discussion by considering 
the $2$--dimensional torus ${\mathcal T}^2=S^1 \times S^1$
(the cartesian product of two unit circles). 
A point in ${\mathcal T}^2$ can be specified by the  
coordinates $(e^{i \theta_1},e^{i \theta_2})$ in
the complex plane and in particular 
a point lying on the longitude has coordinate $(e^{i \theta_1},1)$ while 
a point in a meridian has coordinate $(1,e^{i \theta_2})$. 
 A {\em solid torus} ${\mathcal V}^3$ is a $3$--manifold 
homeomorphic to $S^1 \times {\mathcal D}^2$, where 
${\mathcal D}^2$ is the $2$--dimensional disc. The boundary
$\partial {\mathcal V}^3$ of ${\mathcal V}^3$ is the torus
${\mathcal T}^2$. \\
A {\em framing} on the solid turus is a particular 
homeomorphism $h: S^1 \times {\mathcal D}^2 \rightarrow {\mathcal V}^3$. 
Given a framing $h$ of ${\mathcal V}^3$, $h(1 \times \partial{\mathcal D}^2)$ 
is a meridian while $h(S^1 \times 1)$ is a longitude. \\
A {\em tubular neighborhood} ${\mathcal N}$ of a knot $K$ in 
$S^3$ is an embedding $\iota: K \times {\mathcal D}^2 \rightarrow S^3$ 
such that $\iota (x,0)\,=\,x \; \forall x \in  K$. 
The framing $K^f$ of a knot $K$ 
is defined as the framing of the tubular neighborhood of the 
knot. In particular, the {\em standard framing} $K^f$ of a 
knot $K$ is such that the linking number $\chi(K,K^f)$  is 
equal to zero.\\
 Representing the unit disc ${\mathcal D}^2$ in the complex plane, 
the points of $S^1\times {\mathcal D}^2$ have coordinates 
$(e^{i \theta_1},r e^{i \theta_2})$ with $0\leq r \leq 1$. 
The self--homeomorphisms $\hat{t}$ of $S^1\times {\mathcal D}^2$, 
defined explicitly by 
\begin{equation}
\hat{t}_{\pm}\,(e^{i\theta_1},\,r e^{i \theta_2})\,=\,(e^{i\theta_1},\,r e^{(\theta_2\pm \theta_1)}),
\end{equation}
are the basic twist operations of $S^1\times {\mathcal D}^2$.\\
 Given  a tubular neighborhood ${\mathcal N}$ of a knot 
${\mathcal K}$ with standard framing $f:S^1\times {\mathcal D}^2 \rightarrow {\mathcal N}$, 
the right--handed and left--handed twist $t_{\pm}$ of ${\mathcal N}$ 
are related to the $\hat{t}_{\pm}$ by
\begin{equation}
t_\pm \,=\,f \,\hat{t}_\pm \,f^{-1}.
\end{equation}

A {\em Dehn surgery} performed 
along a (framed) knot $ K \subset S^3$ can be described as follows.
\begin{enumerate}
\item remove the interior $\stackrel{\circ}{{\mathcal N}}$ of a tubular neighborhood 
${\mathcal N}$ of $K$ (the resulting manifold $(S^3 \smallsetminus \stackrel{\circ}{{\mathcal N}})$ 
is the complement torus);
\item consider $(S^3 \smallsetminus \stackrel{\circ}{{\mathcal N}})$ and ${\mathcal N}$ as distinct spaces;
\item glue back ${\mathcal N}$ and $(S^3 \smallsetminus \stackrel{\circ}{{\mathcal N}})$ by 
identifying the points in their boundaries through a given homeomorphism 
$h:\partial{\mathcal N} \rightarrow \partial (S^3 \smallsetminus \stackrel{\circ}{{\mathcal N}})$.
\end{enumerate}
The resulting manifold ${\mathcal M}^3_K$ is recovered by setting
\begin{equation}\label{presM}
{\mathcal M}^3_K\,=\,(S^3 \smallsetminus \stackrel{\circ}{{\mathcal N}})\;\bigcup_h \,{\mathcal N}
\end{equation}
and it is completely specified by the 
knot $K$ and by the chioce of the gluing homomorphism $h$.
Equivalently, the surgery is characterized by the knot $K$ 
and by a closed curve $\gamma \in \partial {\mathcal N}$ representing 
$h(\mu)$, where $\mu$ is the meridian of ${\mathcal N}$. 

\vspace{10pt}

Dehn surgery is a simple and constructive prescription 
which basically consists in removing and 
sewing back solid tori from the $3$--sphere. 
However, since different surgery instructions 
may give rise to homeomorphic manifolds, it is crucial 
to define the equivalence relations that identify the 
surgery instructions providing the same (homeomorphism class of)
$3$--manifold. Once these rules are taken into account, 
the classification problem for  
$3$--manifolds can be actually reduced to the problem of classification of 
knots (links).
 
\begin{figure}[htbp]
	\centering
		\includegraphics[width=0.50\textwidth]{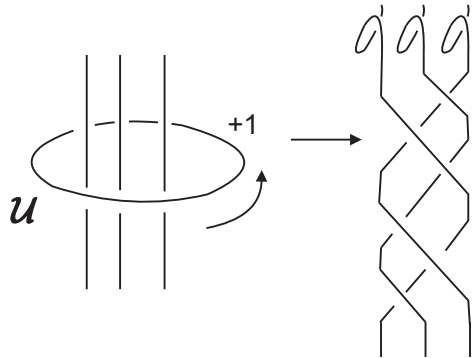}
		\caption{The first Kirby move.}
	\label{fig:kirby1}
\end{figure}

The equivalence relations among 
surgery instructions yielding a same $3$--manifold
are topological operations on framed link diagrams known as 
{\em Kirby moves}.\\ 
{\bf I move}. The configuration described by the unknot 
${\mathcal U}$, with framing $+1$, enclosing 
$n$ unlinked strands $K_i$, with framing $f_i$,
 lying on a ribbon 
can be changed into the configuration where ${\mathcal U}$ 
is removed and the ribbon is twisted in the clockwise 
direction from below, see Fig. [\ref{fig:kirby1}]. The framing 
of the components $ K_i$'s must be changed 
according to 
\begin{equation}
f_i \,\mapsto\; f_i^\prime \,=\,f_i-\chi^2( K_i,\,{\mathcal U}),
\end{equation}
where $\chi^2(K_i,\,{\mathcal U})$ is the 
square of the linking number between $ K_i$ and 
${\mathcal U}$.

\begin{figure}[htbp]
	\centering
		\includegraphics[width=0.50\textwidth]{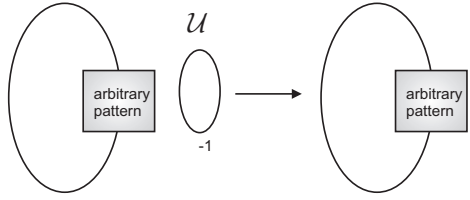}
		\caption{The second Kirby move.}
	\label{fig:kirby2}
\end{figure}

\noindent
{\bf II move}. An unknotted link ${\mathcal U}$ with 
framing $-1$ can be removed without affecting the 
rest of the link, see Fig. [\ref{fig:kirby2}];

\noindent
{\bf III move}. This is the `inverse' of the first 
move and amounts to change the configuration containing ${\mathcal U}$ with 
framing $-1$ into the anti--clockwise twisted ribbon of $K_i$'s
 and without ${\mathcal U}$. Here the framings change according to 
\begin{equation}
f_i\, \mapsto\; f_i^\prime\,=\,f_i+\chi^2(K_i,\,{\mathcal U}).
\end{equation}
\\
\noindent
{\bf IV move}. This is the `inverse' of the second 
move.  An unknotted link ${\mathcal U}$ with 
framing $+1$ can be removed without affecting the 
rest of the link.

\vspace{10pt}

The extension of the results stated  above in order to deal
with $SU(2)_q$--colored surgery links to be associated with
$3$--manifold invariants can be carried
in a quite straightforward way. The construction 
developed in \cite{GaMaRa1,GaMaRa2} and summarized in section 2.3
basically relies on Kaul unitary representation of
(colored) oriented braids on $2n$ strands \cite{Kau1} 
and thus we need preliminarly to
generalize such a representation to
deal with {\em unoriented} braids and links. The goal can be achieved 
by suitably modifying the eigenvalues of the elementary braiding matrices
in the odd--coupled basis $| {\bf p};{\bf r} \rangle^{\,{\bf j}}$ 
introduced in \eqref{oddbas} according to \cite{Kau2}
$$
U(\sigma_{2l+1}\,)\,|{\bf p};{\bf r} \,\rangle^{\,(\dots, j_{2l+1},\,j_{2l+2}, \dots)}
\;=\,\lambda_l \;|{\bf p};{\bf r}\, \rangle^{\,(\dots, j_{2l+1},\,j_{2l+2}, \dots)}
$$
\begin{equation}\label{neweig}
\text{with}\;\;\;
\lambda_l \,\doteq \,\lambda (j_{2l+1},\,j_{2l+2})\,  \equiv \,
\lambda_l (j,\,j')\,=\, (-)^{|j-j^\prime|-l}\,q^{\,\pm (c_j+c_{j^\prime}-c_l)/2}, 
\end{equation}
where $c_j =j(j+1)$ and $c_{j'} =j'(j'+1)$ are the quadratic Casimir
invariants associated with the irreps $j,j'\,$. 
The $\pm 1$ in the exponent of the parameter $q$ refer to left--handed
(respectively, right--handed) half--twists in two parallel strands carrying the coloring
$j,\,j'$ and it can be easily checked that these eigenvalues
do not depend on the orientations but only on the over/under--crossing features 
(the colored strands are ordered from left to right
as happened for oriented braids, see \eqref{colorbr} and Fig. [\ref{fig:colgenerators}]).\\
It is worth noting that the duality matrices \eqref{3nj} --needed
whenever an even braiding $U(\sigma_{2l})$ has
to be applied-- are independent from orientations of the strands,
so that they can be used in the present context with no further modification.

\vspace{10pt}

According to Theorem 1, what we really have to handle are 
colored links in the
vertical framing $\mathbf{f}$ and then 
the effect of adding or deleting a $\pm1$ in the writhe of the link 
must be properly taken into account 
(this operation in the standard framing would not affect 
the topology of the link). Referring in particular
to a $j$--colored framed unknot $\mathcal{U}$, the associated  link invariant 
turns out to be changed into 
\begin{eqnarray}\label{Kmoves}
{\mathit J}\,[\,{\mathcal U};\,j,\,+1;\,q]&\,=\,&q^{c_j}\,{\mathit J}\,[\,{\mathcal U};\,j,\,0;\,q]\nonumber\\
{\mathit J}\,[\,{\mathcal U};j,\,-1;\,q]&\,=\,&q^{-c_j}\,{\mathit J}\,[\,{\mathcal U};\,j,\,0;\,q],
\end{eqnarray}
where $\pm1$ on the left--handed sides
 denote the $\pm1$ vertical framing while in the right--handed sides
there appears the knot invariant  of the unknot in the standard $0$--framing, whose numerical
value is given by the $q$--dimension
$[2j+1]_q$ defined in \eqref{qdim}. The latter  
relations provide in practice  the operatorial 
content of Kirby moves applied to framed colored links. 
Then the requirement of invariance under Kirby moves of the
forthcoming $3$--manifold topological invariants makes it necessary to
compensate the effects of \eqref{Kmoves} by resorting to
properties of the so--called linking matrix.\\ 
For a framed link $[L;\,\mathbf{f}]$ whose components 
$K_1,K_2,\dots,K_S$ have framings $n_1,n_2,\dots,n_S$ respectively,
the linking matrix is a symmetric matrix defined as
\begin{equation}\label{linkma}
{\bf \chi}\,[L;\,\mathbf{f}]\,=\,
\left(
\begin{array}{ccccc}
n_1&\chi (K_1, K_2)&\chi (K_1,K_3)&\cdots & \chi (K_1,K_S)\\
\chi (K_2,K_1)& n_2 &\chi (K_2,K_3)& \cdots & \chi (K_2,K_S)\\
\vdots & \cdots & \cdots & \cdots & \vdots\\
\chi (K_S, K_1)& \cdots & \cdots & \cdots & n_S
\end{array}
\right),
\end{equation}
where $\chi(K_i,K_j)$ is the linking 
number between the component knots $ K_i$ and $K_j$.
The signature of the linking matrix, denoted by $\sigma [L;\,\mathbf{f}]$, 
is the difference between the 
number of positive  and negative eigenvalues of ${\bf \chi}\,[L;\,\mathbf{f}]$.\\
With these preliminary definitions, let us state the following theorem,
the original proof of which can be found in \cite{ReTu,KiMe} (see also \cite{Lic1}).   
\begin{quote}
{\bf Theorem 2.}
For a closed, connected and oriented $3$--manifold $\mathcal{M}^3_L$
obtained by surgery in the $3$--sphere along an unoriented colored framed link
$[L;\,\mathbf{f},\,\mathbf{j}]$ with $S$ link components 
and for any fixed root of unity $q\,=\,e^{\frac{2 \pi i}{k+2}}$ the quantity
\begin{equation}
\label{k3mi.eq}
{\mathcal I}\,[{\mathcal M}^3_L\,;\,{\bf f};\,q]\,= \,
\alpha^{-\sigma [L;\,\mathbf{f}]}\; 
\sum_{\{{\bf j}\}}\, \mu_{j_1}\,\mu_{j_2}\dotsb \mu_{j_S}\; 
\mathit{J}\,[L;\,\mathbf{f},\mathbf{j}\,;\, q] 
\end{equation}
is a topological invariant of the $3$--manifold endowed with the framing assignment
$\mathbf{f}$ .\footnote{This means in practice that, on applying  Kirby moves I-IV 
to $[L;\,\mathbf{f}] \subset S^3$, the value of the invariant does not change, namely depends
only on the homeomorphism class of the $3$--manifold. 
Note however that we are not in the presence of a {\em complete} $3$--manifold invariant
since there exist topologically distinct manifolds with the same ${\mathcal I}$.\\ 
The extension of the theorem to
deal with surgery operations performed on manifolds topologically
different from $S^3$
and to situations in which $q$--deformations of other semisimple Lie groups are involved
can be found in the reference quoted above. All such invariants are collectively  referred to
as $3$--manifold `quantum invariants' or even as  `state sum models' by noticing that an expression
like \eqref{k3mi.eq} can be interpreted as a partition function 
over suitably weighted `states' represented by the link polynomials with different
colorings.}
\end{quote}
Here $\alpha\equiv \exp{\frac{3 \pi i k }{4(k+2)}}$, 
$\mu_j=\sqrt{\frac{2}{k+2}}\sin{\frac{\pi (2j+1)}{k+2}}$, 
$\mathbf{j} \equiv (j_1,j_2,\dots,j_S)$ run over 
$\{ 0,\frac{1}{2},\dots,\frac{k}{2} \}\;$ and the
summation is performed over all admissible colorings. 
$\mathit{J}\,[L;\,\mathbf{f},\mathbf{j}\,;\, q]$ in the previous expression is the 
`unoriented' counterpart of the polynomial 
for the link $L$ with coloring assignment $\mathbf{j}$ on its components
and the summation is performed over all admissible colorings.

It is worth noting that the presentation of the colored links
used both in section 2.2 (see \eqref{colJon}) and in \eqref{neweig} 
is slightly  different from the presentation
given in Theorem 2 above, even
though we keep on using the same notation $\mathbf{j}$ for the colorings.
In the latter case a coloring is assigned to each of the $S$ link components,
while in the former we label the $2n$ strands of the associated braid
with $n$ colors (see {\em e.g.} Fig. [\ref{fig:platting}]).\\ 
However, what really matters is the fact that both
presentations of the link $L$ give rise to the same invariant, 
as could be proved by exploiting the properties
of the $SU(2)_{\,q}$--representation ring. 
From the computational viewpoint this 
twofold choice does not matter 
as well because we could efficiently implement 
the transformation from a given link diagram to any
associated closed braid on a classical computer
(see  {\em e.g.} section 2 of \cite{GaMaRa3} for a discussion
on classical algorithmic questions about braids and links).
Expression  \eqref{k3mi.eq}
makes it manifest the overall dependence on the framing in the factor
$\alpha^{-\sigma [L;\,\mathbf{f}]}$,
but in what follows we are going to express the link
polynomials
$\mathit{J}\,[L;\,\mathbf{f},\mathbf{j}\,;\, q]$ 
as expectation values of Kaul unitary representation \cite{Kau2}
worked out for an unoriented $2n$--strands braid the
plat closure of which gives the link under examination.


\subsection{Quantum algorithm for approximating $3$--manifold invariants}

In this section we shall describe the quantum algorithm 
for computing the colored polynomial
$\mathit{J}\,[L;\,$ $\mathbf{f},\mathbf{j}\,;\, q]$
and the associate invariant ${\mathcal I}\,[{\mathcal M}^3_L\,;\,{\bf f};\,q]$
defined in \eqref{k3mi.eq}. 
This algorithm is an extension of the quantum algorithm 
proposed in \cite{GaMaRa1,GaMaRa2} which efficiently approximated the value of the colored Jones 
polynomials \eqref{colJon}.

 As anticipated in section 1.2 we
need in the present context the notion of {\em additive approximation} 
introduced \cite{BoFrLo} 
(see also \cite{WoYa}).  
Given a normalized function $g(x)$, 
where $x$ denotes an instance of the problem, we have an 
additive approximation of its value for each $x$ if we 
can associate with $g(x)$ a random variable $Z$ such that 
\begin{equation}\label{appinv1}
\text{pr}\, \left\{ \,\left | g(x) - Z \,\right | \leq \,\eta \right \} \,\geq \,3/4 \; , 
\end{equation}
for any $\eta \geq 0$.
Moreover, the time needed to achieve the approximation 
must be polynomial in the size of the problem 
and in ${\eta}^{-1}$.
Then the problem we are interested in can 
be stated formally as follows.

\begin{quote}
{\bf Approximating $3$--manifold invariants.}\\
Given a framed 
link $[L;\,\mathbf{f}]$ with 
component knots $K_1,K_2,\dots,K_S $, framing 
$\mathbf{f}=(n_1,n_2,\dots,n_S)$, a positive integer $k$, and a set of allowed colors 
$\textbf{j}=\{ 0,1/2,\dots,k/2 \}$, 
we want to sample out 
a random variable $Z$ --representing an additive approximation 
of the value of the normalized $3$--manifold invariant
${\mathcal I}\,[{\mathcal M}^3_L\,;\,{\bf f};\,q]$
evaluated at $q=e^{\frac{2 \pi i}{k+2}}\;$\,-- in  such a way that the following 
condition holds true
\begin{equation}\label{appinv2}
\text{pr}\, \left \{ \left |\, 
{\mathcal I}\,[{\mathcal M}^3_L\,;\,{\bf f};\,q]
\,-\,Z \,
  \right |  \, \leq \, \eta \right \} \, \geq \,3/4 \;.
\end{equation}
The size of the problem is expressed in terms of the number of crossings $\kappa$
of the surgery  link $L$ and by the number of strands of the associate braid,  
(as it happened for colored link polynomials), but we shall need to handle properly 
the whole set of allowed colorings $\mathbf{j}$ and
the  framing $\mathbf{f}$ as well.\footnote{From now on we agree 
that the invariant
is normalized by the product of the $q$--dimensions 
associated with the link components, namely by the factor
$\,\prod_{i=1}^{S}[2j_i+1]_q$.
} 
\end{quote}

As anticipated  in section 2.1, the quantum algorithms
for evaluating (additive approximations of) 
topological invariants in the framework
of quantum CSW theory are based on a two--level
procedure outlined already in \cite{GaMaRa2} in connection with colored
link polynomials and improved here for $3$--manifold
invariants. The rationale underlying our procedure is briefly summarized 
below while technical details are developed in the rest of the section.
\begin{itemize}
\item[{\bf (A)}] Within the computational model of the $q$--deformed
spin network recognizer $\mathcal{A}_{\,q}$ (section 2)
both topological and 
field--theoretic data  --encoded into  
a framed colored link $[L;\,\mathbf{f},\mathbf{j}\,]$  
associated with a unitary braiding operator in the Kaul representation \cite{Kau2}-- 
are efficiently processed. The estimate of the time complexity function
(number of computational steps) required to
complete the calculation given in \eqref{stima} still holds true.
\item[{\bf (B)}] By resorting to standard quantum circuit model techniques
and related approximation schemes it can be shown that
\begin{enumerate}
\item[a)] the start state 
$|\psi\rangle^{\mathcal{I}}$ of the recognizer $\mathcal{A}_{\,q}$
needed for processing the $3$--manifold invariant 
${\mathcal I}\,[{\mathcal M}^3_L\,;\,{\bf f};\,q]$ is efficiently 
encoded into a qubit register;
\item[b)] the braiding operator associated with the
framed link $[L;\,\mathbf{f},\mathbf{j}\,]$ --already split into
a sequence of `elementary' braiding and duality transformations
on the basis of the recognizer design--
can be efficiently compiled on a standard quantum circuit applied to 
the start qubit register;
\item[c)] by resorting to the Hadamard test \cite{AhJoLa}
--an efficient sampling procedure which provides the expectation value of an unitary
operator on a (qubit) state-- it is possible to estimate the value of the 
invariant ${\mathcal I}\,[{\mathcal M}^3_L\,;\,{\bf f};\,q]$ from a series of measurements 
on an ancilla qubit coupled to the start state.
\end{enumerate}

This two--level computational process for the approximation of $3$--manifold invariants
in the sense of \eqref{appinv2} is thus efficient
with respect to the standard model of quantum computation
and not simply from the viewpoint of the $q$--deformed automaton model.

\item[{\bf (C)}] The expectation value of the braiding operator evaluated on
the start state and sampled as described above, can in turn be related
to a suitable probability distribution  
on the language(s) recognized by $\mathcal{A}_{\,q}$.
In view of the properties of such distributions 
({\em cfr.} the concluding remarks  of sections 2.1 and 2.2
and appendix B), the whole procedure can be reinterpreted 
by saying that the automaton recognizes the language
of the braid group with a probability distribution given by the
square modulus of the (normalized) invariant ${\mathcal I}\,[{\mathcal M}^3_L\,;\,{\bf f};\,q]$. 
\end{itemize}

\vspace{10pt}

\begin{figure}[htbp]
	\centering
		\includegraphics[width=0.30\textwidth]{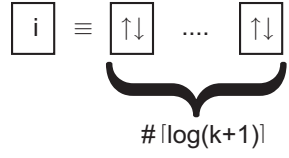}
	\caption{The $i$-th quantum number is encoded into a register made of
	$\lceil \, \log{(k+1)} \, \rceil$ qubits.}
	\label{fig:spincode}
\end{figure}

Given an unoriented link $L$ presented as the plat closure of a $2n$--strands
braid with a fixed set of colorings $\mathbf{j}\equiv
j_1,j_2,\dots,j_{2n}$ (see section 2.2 for the oriented case,
in particular Fig. [\ref{fig:platting}]), let us denote again the conformal block odd--coupled basis
of the boundary WZW theory by $|{\bf p};{\bf r}\,\rangle^{\,\mathbf{j}}$ as in \eqref{oddbas}
and Fig. [\ref{fig:confodd}]. 
Here
${\bf p}= p_0,\dots,p_{n-1} $ and ${\bf r}= r_0,\dots,r_{n-3} $ 
and both $\mathbf{j}$ and the intermediate quantum numbers $ {\bf p},{\bf r} $ 
take values  in the collection $\{ 0,1/2,\dots,k/2 \}$, bounded form 
above by the coupling constant $k$ of quantum CSW theory.\\ 
Each spin quantum number can be encoded into a qubit register made of
$\lceil \,\log_2{(k+1)}\,\rceil$ qubits, where 
$\lceil \, x \, \rceil$ denotes the smallest integer 
$\geq x$, see Fig. [\ref{fig:spincode}].

\begin{figure}[htbp]
	\centering
		\includegraphics[width=0.50\textwidth]{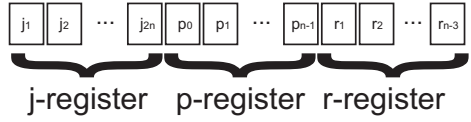}
	\caption{The register of the quantum circuit encoding
     an odd--basis state $|{\bf p};{\bf r}\,\rangle^{\,\mathbf{j}}$.}
	\label{fig:register}
\end{figure}

Since an element of the basis is specified  by 
$(4n-3)$ quantum numbers ({\bf j,p} and {\bf r}), 
we need 
\begin{equation}
\label{qregba}
(4n-3)\times \lceil \,\log_2{(k+1)}\, \rceil
\end{equation} 
qubits to encode  one basis vector. The ordering 
on such quantum register is shown in Fig. [\ref{fig:register}], where
the three sets of quantum numbers are associated with a 
{\bf j}--register, a {\bf p}--register and a {\bf r}--register
respectively. 

\vspace{10pt}

The unitary braiding operator associated with the $2n$--strands unoriented braid 
under consideration can be decomposed --following the scheme explained
in section 2.2 and updating the representation according to \cite{Kau2}-- 
into `elementary' odd braidings and duality trasformation.\\
The elementary braiding matrices are diagonal in the odd--coupled basis, 
so that  their action can be easily implemented on  the quantum register
of Fig. [\ref{fig:register}]. The quantum 
gate realization of $U(\sigma_{2l+1})$  is simply the 
identity matrix on the $({\bf p},{\bf r})$--registers, 
while the ${\bf j}$--register is modified by the action 
of a SWAP gate and a phase gate with a phase factor given by 
the eigenvalue $\lambda (j_{2l+1}, 
j_{2l+2})$ in equation \eqref{neweig}.\footnote{Recall that a SWAP 
acting on two qubits $| x \rangle$, $| y \rangle$ is the operation
$$
\text{SWAP}:| x \rangle | y \rangle \mapsto | y \rangle | x \rangle,
$$
which corresponds to the matrix 
$$
\text{SWAP}=\left( 
\begin{array}{cccc} 
1&0&0&0\\
0&0&1&0\\
0&1&0&0\\
0&0&0&1
 \end{array}
 \right).
$$
}
\begin{figure}[htbp]
	\centering
		\includegraphics[width=6cm]{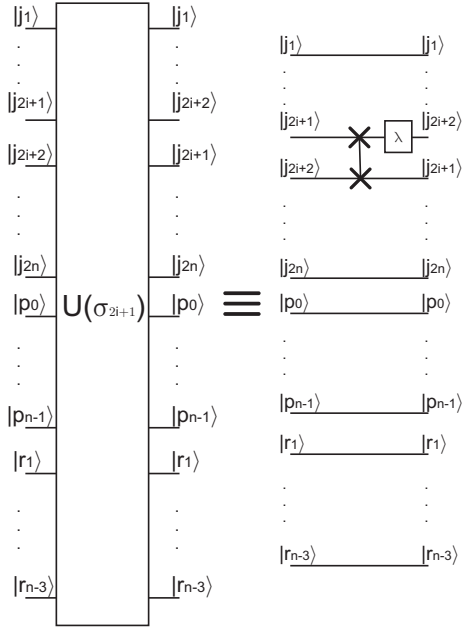}
	\caption{\small{Circuit diagram of the unitary gate associated to an odd braid generator.}}
	\label{fig:oddgate}
\end{figure}

\vspace{5pt}
The duality transformation introduced formally in \eqref{3nj}
must be applied whenever an even elementary braiding matrix
 $U(\sigma_{2l})$ is encountered in order to
recover the current, odd--coupled basis. 
The associated ($q$--deformed)
$3nj$ coefficient can be split into a sequence of
`elementary' duality transformations according to 

 \begin{eqnarray}\label{eq:decomposition}
\nonumber
\lefteqn
{A_{({\bf p};{\bf r})}^{({\bf q};{\bf s})} 
\left[ 
\begin{array}{cc}
j_1&j_2\\ 
j_3&j_4\\ 
\vdots & \vdots \\ 
j_{2n-1}&j_{2n}
\end{array}
 \right] =} \\ 
\nonumber
&& \sum_{t_1 \dots t_{n-2}} \prod_{i=1}^{n-2} 
\left( 
A_{t_i p_i} \left[ \begin{array}{cc} r_{i-1}&j_{2i+1}\\j_{2i+2}&r_i \end{array} \right] 
A_{t_is_{i-1}} \left[ \begin{array}{cc} t_{i-1}&q_i\\s_i&j_{2n} \end{array} \right]
 \right) \\ 
 & \times & \prod_{l=0}^{n-2} A_{r_lq_{l+1}} \left[ 
 \begin{array}{cc} t_l&j_{2l+2}\\j_{2l+3}&t_{l+1} \end{array}
  \right]
\end{eqnarray}
(see Fig. [\ref{fig:decomposition}]
of section 2.2 for the graphical representation of such decomposition).\\
Each symbol in the latter expression is the matrix form of a $q$--$6j$ 
coefficient (see Fig. [\ref{fig:duality2}]), namely
\begin{equation}
\label{eq:6jprima} 
A_{j_{23}}^{j_{12}} \left[ 
\begin{tabular}{cc}
$j_1$ & $j_2$ \\
$j_3$ & $j$
\end{tabular}
\right] \doteq
( - )^{( j_1+j_2+j_3+j)} \,
([ 2j_{12}+1]_q  [ 2_{23}+1 ]_q)^{1/2} 
\left\{
\begin{tabular}{ccc}
$j_1$ & $j_2$ & $j_{12}$ \\
$j_3$ & $j$ & $j_{23}$
\end{tabular}
 \right\}_q\,,
\end{equation}
where the labels of the quantum numbers has been slightly changed
to comply with the following standard explicit expression of the $q$--$6j$ 
\begin{eqnarray} 
\label{eq:6jseconda}
\nonumber
\lefteqn{\left\{\begin{array}{ccc} j_1&j_2&j_{12}\\ j_3&j&j_{23} \end{array} \right\}_q = } \\
 \nonumber
 && {} \Delta(j_1,j_2,j_{12}) \Delta(j_3,j,j_{12}) \Delta(j_1,j,j_{23}) \Delta(j_2,j_3,j_{23}) \\
 \nonumber
 && {} \times \sum_{z \geq 0}
\left[ \frac{(-)^z [z+1]_q!}{[z-j_1-j_2-j_{12}]_q! [z-j_3-j-j_{12}]_q! [z-j_1-j-j_3]_q!} \right. \\
 \nonumber
 && {} \times  \frac{1}{[z-j_2-j_3-j_{23}]_q! [j_1+j_2+j_3+j-z]_q!} \\
 \nonumber
 && \left. {} \times  \frac{1}{[j_1+j_3+j_{12}+j_{23}-z]_q! [j_2+j+j_{12}+j_{23}-z]_q!} \right].\\
\end{eqnarray}
The symbols $\Delta (\dots)$ are combinatorial factors involving
the factorials of $q$--dimensions (see appendix I of \cite{Kau1}),
where the factorial of a $q$--number $[x]_q$ is defined as
$[x]_q!=[x]_q [x-1]_q \ldots [2]_q[1]_q$.\\ 
For each choice of the entries (in the allowed set 
$\{ 0,1/2,\dots,k/2 \}$)
the latter power series is actually a summation over the finite set of (integer and half--integers)
$z$'s that yields non--negative quantum integers. 

\vspace{5pt}

From the remarks above it should be clear that the problem of efficiently 
implementing \eqref{eq:decomposition}, namely the most general change of basis,  
is equivalent to the simpler problem of efficiently 
compiling a sequence of $q$--$6j$ symbols or elementary duality matrices
\eqref{eq:6jprima}. But any such coefficient ({\em i.e.} the symbol 
with a fixed set of entries) can be easily and efficiently evaluated with 
a classical computer owing to the finiteness of its explicit expression 
\eqref{eq:6jseconda}.\\ 
On the other hand,  it is necessary to make explicit the
action of the matrix \eqref{eq:6jprima} on the qubit register, namely
on an Hilbert space of dimension
$2^{\lceil \log{(k+1)}\rceil}$ for each admissible set of entries.
\footnote{The `classical' $6j$--symbol can be expressed in terms of 
a series similar to \eqref{eq:6jseconda}, but the encoding
of such symbols on qubit registers would depend explicitly on the values of the entries,
so that the computation is not obvioulsy efficient with respect
to the size of the input data, see footnote 2 in section 2.1.} 

\begin{figure}[htbp]
	\centering
		\includegraphics[width=0.50\textwidth]{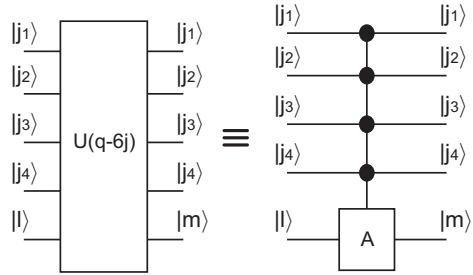}
	\caption{The multiplexor gate representing the action of 
	a $q$--$6j$ transformation.} 
	\label{fig:dualitygate}
\end{figure}

\begin{figure}[htbp]
	\centering
		\includegraphics[width=0.50\textwidth]{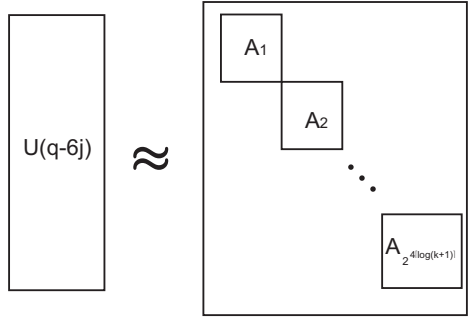}
	\caption{The block--decomposition of the matrix corresponding to 
	a $q$--$6j$ transformation.}
	\label{fig:dgstructure}
\end{figure}

The circuit realization is shown in Fig. [\ref{fig:dualitygate}], where 
the {\bf j}--register 
acts as a control register on the qubits of the {\bf p}({\bf r})--registers 
involved in the transformation. The transformation $A$
can then be thought of as a  gate with a `block structure', or a
multiplexor associated with the block structure depicted in Fig. [\ref{fig:dgstructure}]. 
Each block 
corresponds to a particular configuration of the {\bf j}--qubits, and 
the matrix element inside the block are $q$--$6j$, up to suitable factors 
(equation \eqref{eq:6jprima}). 

On the basis of the decomposition in \eqref{eq:decomposition} we realize that the
allowed elementary duality matrices are always parametrized 
by the set ${\bf j}$ of those quantum numbers which remain 
unchanged when they are applied to the proper registers. 
The crucial remark here consists in noticing that the dimension of these matrices is independent 
of the size of the problem, determined by the index of the braid group 
and the number of crossings. Since there exist efficient 
methods to approximate unitary matrices of a given dimension \cite{ShBuMa}, 
a sequence of universal gates can be always worked out that 
efficiently approximate every $q$--$6j$ as well
(see \cite{Sil} for more details).

\begin{figure}[htbp]
	\centering
		\includegraphics[width=0.50\textwidth]{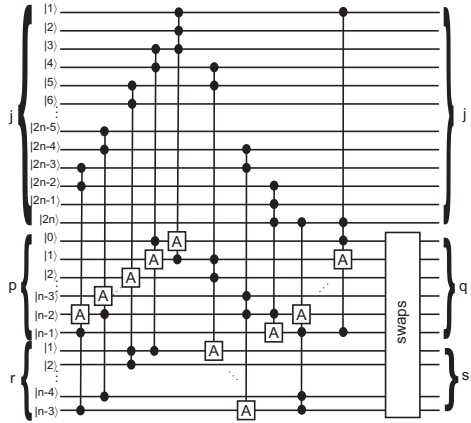}
	\caption{The circuit realization of a general $q$--$3nj$ recoupling transformation.}
	\label{fig:general3nj}
\end{figure}

\begin{figure}[htbp]
	\centering
		\includegraphics[width=0.50\textwidth]{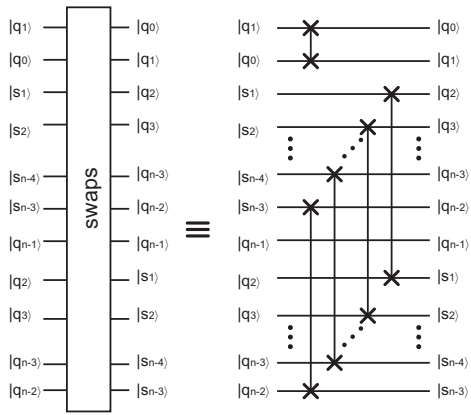}
	\caption{The sequence of SWAP gates needed to reordering the qubit register.}
	\label{fig:extswap}
\end{figure}

The number of elementary 
duality transformations needed to decompose 
a general duality transformation \eqref{fig:decomposition}
is $(2n-3)$, linear in the size of the problem under consideration.
The action of a  $q$--$3nj$ recoupling transformation on 
the register of Fig. [\ref{fig:register}] 
is shown in Fig. [\ref{fig:general3nj}].
Note however that, once all the $(2n-3)$ gates represented by $q$--$6j$  
have been applied, it is necessary to swap some of the qubits 
in order to recover the proper order in the register,  see Fig. [\ref{fig:extswap}].

\begin{figure}[htbp]
	\centering
		\includegraphics[width=0.50\textwidth]{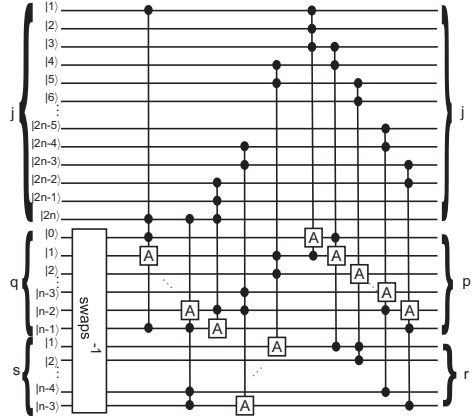}
	\caption{The inverse of the circuit of Fig. [\ref{fig:general3nj}].}
	\label{fig:inverse3nj}
\end{figure}

Upon applying a $q$--$3nj$ recoupling transformation we end up in
 the even--coupled basis $|{\bf q};{\bf s}\rangle^{\,\mathbf{j}}\,$ 
(see   \eqref{evebas} and Fig. [\ref{fig:confeven}]) which 
diagonalizes the even braiding matrices $U(\sigma_{2l})$
of the Kaul representation. Their action on the (even) qubit register can then be implemented
by paralleling the procedure described in the odd case.\\
The further step consists in going back to the $|{\bf p};{\bf r}\rangle^{\,\mathbf{j}\,}$ 
basis, which can be achieved by means of the finite sequence of 
$q$--$6j$ transformations shown in Fig. [\ref{fig:inverse3nj}],
eventually completed by a suitable sequence of SWAPS to recover the
initial ordering of the start register.

This completes the analysis of the quantum circuit which implements
efficiently the Kaul representation \cite{Kau2} on the Hilbert space 
spanned by the odd--coupled basis of conformal blocks
(point b) at the beginning of this section).

\vspace{5pt}

For what concerns the preparation of the start state 
according to point a), notice once more that
we have considered so far the case of an unoriented link $[L;\mathbf{j}]$
(plat closure of a $2n$--strands braid) with a fixed coloring set $\mathbf{j}$.
As already pointed out in the case of oriented links (see section 2.2, \eqref{colJon}
and appendix B) the start state of the automaton calculation 
is a singlet vector in the odd--coupled basis   
\begin{equation}\label{unLst}
[L;\mathbf{j}\,]\;\;\;\leftrightarrow \;\;
|{\bf 0};{\bf 0}\,\rangle^{\,\mathbf{j}}\,;\;\;
{\bf p}= p_0,\dots,p_{n-1}\equiv {\bf 0}\,,\; {\bf r}= r_0,\dots,r_{n-3} \equiv {\bf 0},
\end{equation}
which can be efficiently encoded into the qubit register of 
Fig. [\ref{fig:register}] as a particular case of the construction
carried out above. Then the polynomial for an unoriented, framed link
$\mathit{J}\,[L;\,\mathbf{f},\mathbf{j}\,;q\,]$ can be recovered 
as the (trace of the) expectation value of the associated braiding
operator in the updated Kaul representation.
However, the start state 
$|\psi\rangle^{\,\mathcal{I}}$ of the recognizer $\mathcal{A}_{\,q}$
needed for processing the $3$--manifold invariant 
${\mathcal I}\,[{\mathcal M}^3_L\,;\,{\bf f};\,q]$ 
must be prepared in order to comply with the
summation over the colorings  in \eqref{k3mi.eq}. 
The goal can be achieved in a quite staightforward way
by resorting to quantum parallelism,
as described in the following.

As before,  we first split the qubit register into three ordered parts, the 
{\bf j}-, {\bf p}-- and {\bf r}--registers. 
Then we initialize 
the {\bf j}--register into the following weighted superposition 
of states
\begin{equation}\label{superp}
|\psi\rangle^{J}\doteq \sum_{\{ {\bf j}\}}\,\sqrt{\mu_{{\bf j}}}\;|{\bf j}\,\rangle,
\end{equation}
with
$\mu_{{\bf j}}\equiv \mu_{j_1}\dotsb \mu_{j_S}$
($S$ in the number of link components and the coefficients $\mu$'s
are defined as in Theorem 2 of section 3.1). 
Since the link $L$ can be considered as  the plat closure of 
a braid word $w$, the quantum circuit needed to encode
the start state can be implemented  by the following composition of 
unitary gates  
\begin{equation}\label{supgate}
S^J(w)\,U^{J \otimes P \otimes R}(w): | \psi \rangle^J \otimes | {\bf 0};{\bf 0} \rangle 
 \mapsto |\psi\rangle^J \otimes | {\bf p}^\prime;{\bf r}^\prime \rangle^{P \otimes R}, 
\end{equation}
where $S^J(w)$ is the SWAP gate on the {\bf j}--register and 
 $| {\bf p}^\prime;{\bf r}^\prime \rangle$ is the state 
obtained from the application  of the circuit $U(w)$.

The following sequence of 
equalities shows that this quantum circuit actually gives
the required $3$--manifold invariant

\begin{eqnarray}
\label{gateinv}
\lefteqn{^{P\otimes R}\!\langle {\bf 0};{\bf 0} |\otimes 
\, ^{J}\!\langle \psi | S^J(w) U^{J\otimes P\otimes R}(w) | \psi \rangle^J \, 
| {\bf 0};{\bf 0} \rangle^{P\otimes R}=}\nonumber\\
&&= ^{P\otimes R}\!\langle {\bf 0};{\bf 0} |\otimes \,
^{J}\!\langle \psi | \cdot | \psi \rangle^J \otimes U^{P\otimes R}(w)|{\bf 0};{\bf 0} 
\rangle^{P\otimes R}=\nonumber\\
&&= \sum_{\{{\bf j}\}} \mu_{{\bf j}} \,\langle {\bf 0};{\bf 0} | U^{P \otimes R}(w) | {\bf 0};
{\bf 0} \rangle=\nonumber\\
&&={\mathcal I}[\mathcal{M}^3_L\,;{\bf f}\,;q] \,\alpha^{\sigma[L;\,\mathbf{f}]}.
\end{eqnarray}
Recall that the signature of the linking matrix
$\sigma[L;\,\mathbf{f}]$ can be easily 
computed classically once  the linking matrix \eqref{linkma}
(which is part of the topological input data of the
algorithmic problem) is given. 

\vspace{10pt}

Finally, the efficient sampling of the value of the invariant 
referred to in point c) can be  carried out by resorting to the  techniques
of \cite{AhJoLa,WoYa}, as already done in \cite{GaMaRa2} for the case of colored link 
polynomials.


\section{Concluding remarks}

We have shown that all the significant quantities --partition functions and
observables-- in $SU(2)$ quantum CSW theory can be efficiently approximated
by quantum algorithms  at finite values of the
coupling constant $k$, reflecting  the intrinsic field--theoretic solvability of this theory.
The efficiency of the quantum algorithms is proved on the basis of a two--level
computational scheme which relies on the implementation of unitary representations of the braid group 
proposed in \cite{Kau1,Kau2}. In particular, as shown in section 3.2, the representation needed
to handle $3$--manifold quantum invariants can be efficiently compiled on
a quantum circuit equipped with a suitable start qubit register,
namely within the `standard' model of quantum computation.\\
The relevance of our result in connection with the algorithmic complexity of quantum CSW
theory has been extensively addressed in section 1.2, while further developments and
applications to geometry and quantum gravity models had  already been discussed in 1.3.\\
More generally, it would be interesting to improve our approach, on the one hand,
by handling quantum topological invariants arising from CSW theories 
with arbitrary semi--simple Lie
groups \cite{KaRa} and, on the other,
by exploring the quantum computational complexity
of solvable models in statistical mechanics \cite{Lid}.

\vspace{5pt}

Finally, let us comment in some more details 
the issue concerning the model of quantum computation adopted .
As pointed out in {\bf (A)} and {\bf (C)} of section 3.2, 
a central role in our procedure is
played by the quantum recognizer 
$\mathcal{A}_{\,q}$ able to process efficiently
the language generate by the braid group with
transition matrices given by Kaul unitary representations
(section 2.2 and 3.2) and probability distributions associated with 
quantum topological invariants. It may be tempting to proceed without
this step, processing directly  the unitary representations 
within the quantum circuit scheme of computation. 
However,  we proved in section 3.2
that the basic morphisms of the $q$--tensor category 
$(\,\mathfrak{R} (SU(2)_q\,)\,;\,\mathcal{R}\,;\,\mathcal{F}\,)$
on which the recognizer is modelled can be efficiently compiled
and approximated on a quantum circuit 
(in particular the implementation of a $q$--$6j$ transformation
is independent of both the input size of the algorithmic problem and
on the values of its entries; see also footnotes  2, section 2.1 and 8, section 3.2).

This achievement, quite remarkable as it is by itself,  opens as well 
the further  possibility of looking  at the $q$--spin network simulator
as the fundamental model of  computation  for a wide range of
algorithmic problems in geometric topology and group theory. 
According to the quantum recognizer definition given
in section 2.1, the specific problems that can be dealt with it require 
the selection of suitable start states
--that must be efficiently encoded into qubit registers-- and of
(possibly constrained) sets of transition matrices. We conjecture 
that the approximate evaluation of invariants for (colored) triangulations in dimension
$2$ and $3$ should require  minor modifications of the
scheme employed in this paper for processing quantum invariants of links and $3$--manifolds.
  
 \vspace{20pt}

\noindent \textbf{Acknowledgments}\\
We are in debt with Romesh Kaul for clarifying remarks on his work
and for suggesting us possible further developments.

\vfill
\newpage
\section*{Appendix A. The spin network quantum simulator}

The universal model of quantum computation proposed in
\cite{MaRa1,MaRa2,MaRa3} relies on the (re)coupling theory of $SU(2)$ angular momenta 
\cite{BiLo9,Russi}. 
It can be thought of as a generalization of the standard 
quantum circuit model in which  the computational Hilbert spaces
are binary--coupled eigestates of $N\equiv(n+1)$ $SU(2)$--angular momenta
(whose quantum numbers range over $\{0,1/2,1,3/2,\ldots \}$)
and unitary transformations (`gates') are  expressed in terms of
recoupling coefficients ($3nj$ symbols) connecting pairs of 
inequivalent binary coupling schemes.\footnote{The model can be 
extended to include Wigner rotations in the eigenspace of
the total angular momentum, see section 3.2 of \cite{MaRa2}.}

The architecture of the spin network is modelled as an $SU(2)$ fiber space structure
over a discrete base space $V$
\begin{equation}\label{1}
(V,\,\mathbb{C}^{2J+1}\,)_n
\end{equation}
which encodes all possible computational Hilbert spaces as well
as all gates for any fixed number $N=n+1$ of incoming angular momenta
(see appendix A of \cite{MaRa2} and \cite{MaRa3} for more details).
The base space $V\;\doteq\;\{v(\mathfrak{b})\}$ represents the vertex set of a regular,
$3$--valent graph, the so--called twist--rotation graph \cite{Belgi}
$\mathfrak{G}_n(V, E)$ with cardinality $|V| = (2n)!/n!$, {\em i.e.}
the quadruple factorial number. 
$E$, the edge set of the graph, will be associated with permitted transformations
between pairs of verices as described below.\\
There exists a 1:1 correspondence
\begin{equation}\label{2}
\{v(\mathfrak{b})\}  \longleftrightarrow \{\mathcal{H}^J_n\,(\mathfrak{b})\}
\end{equation}
between the vertices of $\mathfrak{G}_n(V, E)$ and the computational Hilbert spaces 
of the simulator, where $\mathfrak{b}$
denotes the binary bracketing structure that we are going to describe. 
For a given value of $n$, $\mathcal{H}^J_n(\mathfrak{b})$ is the simultaneous
eigenspace of the squares of $2n+1$ Hermitean, mutually commuting angular
momentum operators
$${\bf J}_1,\;{\bf J}_2,\;{\bf J}_3,\ldots,{\bf J}_{n+1}\,\equiv \,\{{\bf J}_i\};\;\;\;\; 
{\bf J}_1\,+\,{\bf J}_2\,+\,{\bf J}_3\,+\ldots+{\bf J}_{n+1}\;\doteq\;{\bf J}$$
\begin{equation}\label{3}
\text{and}\;\;\;\;{\bf K}_1,\,{\bf K}_2,\,{\bf K}_3,\,\ldots,\,{\bf K}_{n-1}\,\equiv \,\{{\bf K}_h\}
\end{equation}
together with the operator $J_z$ (the projection of the total angular momentum $\bf{J}$
along the quantization axis). 
The associated quantum numbers are 
$j_1, j_2,\ldots,j_{n+1}$ $;\,J;$ $ k_1,k_2,\ldots,$ $k_{n-1}$ and $M$, where $-J \leq M
\leq$ in integer steps. If
${\cal H}^{j_1}\otimes$ ${\cal H}^{j_2}\otimes\cdots$ $\otimes 
{\cal H}^{j_{n}}\otimes {\cal H}^{j_{n+1}}$
denotes the factorized Hilbert space, namely the $(n+1)$--fold tensor product 
of the individual eigenspaces of the $({\bf J}_i)^2\,$'s, the operators 
${\bf K}_h$'s represent intermediate angular momenta generated, through Clebsch--Gordan series, 
whenever a pair of ${\bf J}_i$'s are coupled.\\
 As an example, by coupling
sequentially the ${\bf J}_i$'s according to the scheme
$(\cdots(({\bf J}_1+{\bf J}_2)+{\bf J}_3)+\cdots+{\bf J}_{n+1})$ $={\bf J}$ -- which generates
$({\bf J}_1+{\bf J}_2)={\bf K}_1$,
$({\bf K}_1+{\bf J}_3)={\bf K}_2$, and so on --
we would get a binary bracketing structure of the type
$(\cdots((({\cal H}^{j_1}\otimes{\cal H}^{j_2})_{k_1}$ $\otimes{\cal H}^{j_3})_{k_2}
\otimes$ $\cdots \otimes
{\cal H}^{j_{n+1}})_{k_{n-1}})_J$, where we add an overall  bracket labeled by the quantum
number of the total angular momentum $J$. Note that, as far as $j_i$'s
 quantum numbers are involved, any value belonging to 
 $\{0,1/2,1,3/2,\ldots \}$ is allowed, while the ranges of the $k_h$'s are suitably 
 constrained by Clebsch--Gordan decompositions
 ({\em e.g.}, if $({\bf J}_1+{\bf J}_2)={\bf K}_1$ $\Rightarrow$ $|j_1-j_2| \leq$
 $k_1 \leq j_1+j_2$).
We denote a binary coupled basis of $(n+1)$ angular
momenta in the $JM$--representation 
and the corresponding Hilbert space appearing in (\ref{2}) as
$$\{\,|\,[j_1,\,j_2,\,j_3,\ldots,j_{n+1}]^{\mathfrak{b}}\, ;k_1^{\mathfrak{b}\,},\,k_2^{\mathfrak{b}\,}
,\ldots,k_{n-1}^{\mathfrak{b}}\, ;\,JM\, \rangle,\;
-J\leq M\leq J \}$$
\begin{equation}\label{4}
=\;{\cal H}^{J}_{\,n}\;(\mathfrak{b})\;\doteq\;\mbox{span}\;\{\;|\,\mathfrak{b}\,;JM\,\rangle_n\,\}\;,
\end{equation}
where  the string inside $[j_1,\,j_2,\,j_3,\ldots,j_{n+1}]^{\mathfrak{b}\,}$ is not necessarily
ordered, $\mathfrak{b}$ is the shorthand notation for the current binary bracketing structure and 
the $k_h$'s are uniquely associated with the chain of pairwise couplings selected by $\mathfrak{b}$.\\
For a given value of $J$
each $\mathcal{H}^J_n (\mathfrak{b})$ has dimension $(2J + 1)$ over 
$\mathbb{C}$ and thus there exists one isomorphism
\begin{equation}\label{5}
\mathcal{H}^J_n (\mathfrak{b})\;\;\; \cong _{\,\mathfrak{b}}\;\;\; \mathbb{C}^{2J+1}
\end{equation}
for each admissible binary coupling scheme $\mathfrak{b}$ of $(n + 1)$ incoming spins.
It is worth stressing that such isomorphic spaces are physically inequivalent
as far as they are associated with different schemes of (binary) interactions.\\
The vector space $\mathbb{C}^{2J+1}$ is interpreted as the typical fiber attached to each vertex
$v(\mathfrak{b}) \in V$ of the fiber space structure (\ref{1}) through the isomorphism (\ref{5}).

For what concerns unitary operations acting on the computational
Hilbert spaces (\ref{4}), it can be shown \cite{BiLo9} 
that any $3nj$ symbols of 
$SU(2)$ can be 
splitted into `elementary gates' represented by Racah and phase transformations
(in the categorical language of section 2.1 they are referred to as the basic morphisms introduced
in \eqref{assoc} and \eqref{twist}, respectively).
A Racah transform applied to a basis vector of the type \eqref{4} is defined formally as
\begin{equation}\label{6}
F\;:\,\;| \dots (\,( a\,b)_d \,c)_f \dots;JM \rangle\;\, \mapsto \;\,
\,|\dots( a\,(b\,c)_e\,)_f \dots;JM \rangle, 
\end{equation}
where we are using here Latin letters $a,b,c,\ldots$ to denote both incoming ($j_i\,$'s
in the previous notation) 
and intermediate ($k_h\,$'s) spin quantum numbers.
The explicit expression of (\ref{6}) reads
$$|(a\,(b\,c)_e\,)_f\,;M\rangle$$
\begin{equation}\label{7} 
=\sum_{d}\,(-1)^{a+b+c+f}\; [(2d+1)
(2e+1)]^{1/2}
\left\{ \begin{array}{ccc}
a & b & d\\
c & f & e
\end{array}\right\}\;|(\,(a\,b)_d \,c)_f \,;M\rangle,
\end{equation}
where there appears the Racah--Wigner $6j$ symbol of $SU(2)$ and $f$ here plays the role
of the total angular momentum quantum number. Owing to the 
Wigner--Eckart theorem, the magnetic quantum number is not affected by
such transformation, and the same holds true for a general, $3nj$ recoupling
coefficient. Recall also that 
the square of the $6j$ symbol in (\ref{7}) represents the 
probability that a system prepared in the state $|(\,(a\,b)_d \,c)_f \,;M\rangle$
will be measured in the state $|(a\,(b\,c)_e\,)_f\,;M\rangle$.

A phase  transform on  a basis vector \eqref{4} is defined as
\begin{equation}\label{8}
R\;:\,\;| \dots ( a\,b)_c  \dots;JM \rangle\;\, \mapsto \;\,
\,|\dots \,(b\,a)_c\, \dots;JM \rangle, 
\end{equation}
and esplicitly reads 
\begin{equation}\label{9}
| \dots ( a\,b)_c  \dots;JM \rangle \,=\,
(-)^{a+b-c}\;|\dots \,(b\,a)_c\, \dots;JM \rangle, 
\end{equation}

\begin{figure}[htbp]
	\centering
		\includegraphics[width=0.60\textwidth]{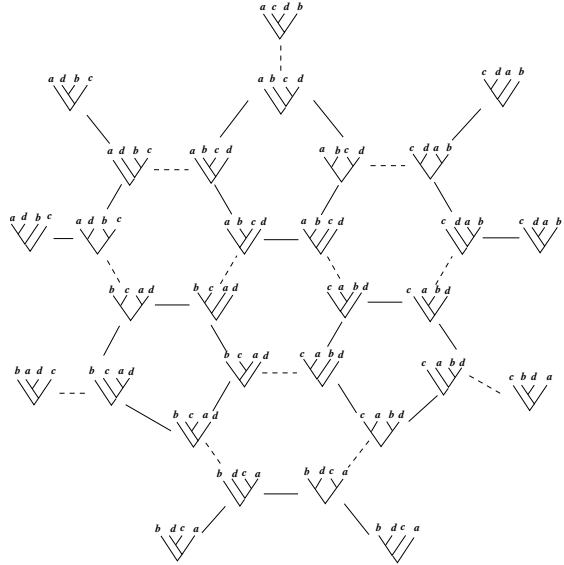}
	\caption{A portion of the spin network graph $\mathfrak{G}_3(V, E)$. The vertices
     are labelled by rooted binary trees encoding the combinatorics
     of the binary coupled computational Hilbert spaces
      while the  edges represent Racah transformations and the dashed ones are phases.}
	\label{fig:TRgraph}
\end{figure}

The edge set $E = \{e\}$ of the twist--rotation graph
$\mathfrak{G}_n(V, E)$ is a subset of the Cartesian
product $(V \times V )$ where 
an (undirected) arc between two vertices $v(\mathfrak{b})$ and $v(\mathfrak{b}')$
\begin{equation}\label{10}
e\,(\mathfrak{b},\mathfrak{b}')\;\doteq \;(v(\mathfrak{b}),\, v(\mathfrak{b}')) 
\;\in \;(V \times V)
\end{equation}
exists iff the underlying Hilbert spaces are related to each other by 
an elementary unitary operation of the type (\ref{6}) or \eqref{8}. 
Note also that elements in $E$ can be considered as mappings
$$
(V\,\times\,\mathbb{C}^{2J+1})_n\; \longrightarrow\,
(V\,\times\,\mathbb{C}^{2J+1})_n
$$
\begin{equation}\label{11}
\;\;\;\;\;\;\;(v(\mathfrak{b}),\,\mathcal{H}^J_n (\mathfrak{b})\,)\, \mapsto\,
(v(\mathfrak{b}'),\,\mathcal{H}^J_n (\mathfrak{b}')\,)
\end{equation}
connecting each given decorated vertex to one of its nearest 
vertices and thus define a `transport 
prescription in the horizontal sections' belonging to the total space
$(V \times \mathbb{C}^{2J+1})_n$ of the fiber bundle (\ref{1}). \\
The structure of the graph $\mathfrak{G}_n(V, E)$ in the case of $(n+1)=4$ incoming
spin variables $a,b,c,d$ is shown in Fig. [\ref{fig:TRgraph}] 
(such a combinatorial pattern encoding both binary--coupled Hilbert spaces 
and trasformations among them was used for the first time in \cite{AqCo}).

The crucial feature that characterizes the graph $\mathfrak{G}_n(V, E)$ arises from 
the compatibility conditions relating the basic morphisms \eqref{6} and \eqref{8}
(referred to as the exagon and pentagon relations, see section 2.1).
The Racah identity and the Biedenharn--Elliott identity
together with the orthogonality conditions for $6j$ symbols (see {\em e.g.} \cite{Russi}
for their explicit expressions)
ensure that any simple path in $\mathfrak{G}_n(V, E)$ with fixed endpoints can 
be freely deformed into any other, providing identical quantum transition amplitudes
at the kinematical level.

\vspace{10pt}

\begin{figure}[htbp]
\begin{center}
\includegraphics[width=6cm]{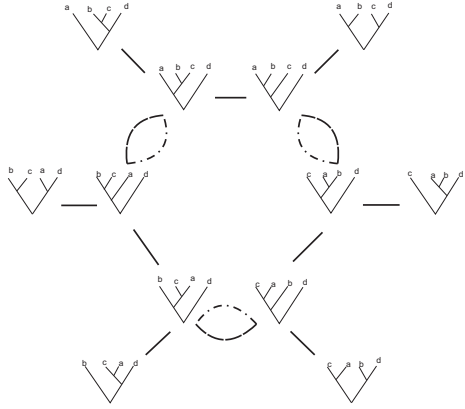}
\end{center}
\caption{A portion of the $q$--deformed twist--rotation graph $(\mathfrak{G}_3 (V, E))_q$:
with respect to the previous figure, each phase trasformation 
has been splitted in order to make manifest the non trivial braiding features.}
\label{fig:brTRgraph}
\end{figure}

For what concerns the $q$--deformed spin network  modelled on the
$q$--tensor category $(\,\mathfrak{R} (SU(2)_q\,)\,;\,\mathcal{R}\,;\,\mathcal{F}\,)$
defined in \eqref{qcategory} of section 2.1, we omit here all
techical details and refer the reader to section 4 of \cite{GaMaRa2} (see also the appendix
of \cite{Kau1} for both definitions and notations). As already pointed out, the basic morphism $\mathcal{F}$
is implemented in this case by means of a $q$--$6j$ symbol defined in
\eqref{eq:6jseconda} while the braiding
$\mathcal{R}$ is to be associated with the over-- or under--crossings
of two contiguous strands belonging to  (colored) braids 
(see {\em e.g.} the eigenvalue equation \eqref{neweig} of the odd braiding 
operators employed in section 3.2).\\
 A pictorial  representation
of $q$--deformed spin network graph is shown in Fig. [\ref{fig:brTRgraph}].

\section*{Appendix B. Automaton calculation}

\begin{figure}[htbp]
	\centering
		\includegraphics[width=0.40\textwidth]{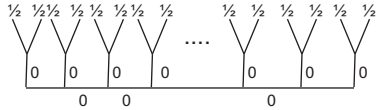}
\caption{The start state of the $q$--spin network automaton with 
	probability distribution associated with the Jones polynomial.}
	\label{fig:instate}
\end{figure}

The Jones polynomial \cite{Jon}
is a particular instance of colored link polynomial where
the labels of the 
conformal block basis correspond to the fundamental, $\tfrac{1}{2}$--irrep 
of $SU(2)_q$. This allows us to simplify the notation 
for the states of the recognizer $\mathcal{A}_{q\,}$ defined in section 2.2
by setting 
$$
| {\bf p};{\bf r} \rangle^{\bf j} \rightarrow | {\bf p};{\bf r} \rangle,
$$
since the coloring assignment $\{ {\bf j} \}$ is always a string of $2n$
$\tfrac{1}{2}$--spins.
The start state of the automaton is chosen  to be $| {\bf 0};{\bf 0} \rangle$, 
{\em i.e.} all the internal labels are equal to $0$, as depicted in Fig. 
[\ref{fig:instate}]. 

As discussed at the end of section 2.2, let us provide the automaton 
with an input word $w$ of length $\kappa$ corresponding 
to an  unitary evolution $U(w)$ expressed in terms of the sequence
given in \eqref{brevol}.

In order to complete the definition of $\mathcal{A}_{\,q}$ 
 we need explicit expressions 
for $P$ (accept) and $P$ (reject) (see section 2.1). We choose the 
following
\begin{equation}\label{prJ1}
P(\text{accept})\;\equiv \; | {\bf 0};{\bf 0} \rangle \langle {\bf 0};{\bf 0} |\,,
\end{equation}
$$
P(\text{reject})\; \equiv \; {\mathbb I}-| {\bf 0};{\bf 0} \rangle \langle {\bf 0};{\bf 0} |\,.
$$
The `a priori' probability distribution 
for the language generated by the braid group that we choose is the
square modulus of the Jones polynomial of the plat closure $\hat{w}$ 
of the braid $w$, namely
\begin{equation}\label{prJ2}
{\rm Pr}(w)\,=\,|\mathit{J}(\hat{w};\,q)|^2 \,\equiv\, V(\hat{w};\,q)\,,
\end{equation}
where $q$ is the root of unity 
at which  the polynomial is evaluated. 
Using the properties of the Kaul representation
(see \eqref{colJon}) and \eqref{prJ1}
there follows that
\begin{eqnarray*}
\lefteqn{\left| {\rm Pr}(w)-\langle {\bf 0};{\bf 0} | U^\dagger(w) 
P(accept) U(w) | {\bf 0};{\bf 0} \rangle \right|=}\\ 
& & \left| {\rm Pr}(w)-\langle {\bf 0};{\bf 0} | U^\dagger(w) 
| {\bf 0};{\bf 0} \rangle \langle {\bf 0};{\bf 0} |
 U(w) | {\bf 0};{\bf 0} \rangle \right|=\\
 & & \left| {\rm Pr}(w)-\left|\langle {\bf 0};{\bf 0} |
 U(w) | {\bf 0};{\bf 0} \rangle \right|^2 \right|=\\
& & \left| V(\hat{w},q)-\left|\langle {\bf 0};{\bf 0} |
 U(w) | {\bf 0};{\bf 0} \rangle \right|^2 \right|=0\,.
\end{eqnarray*}
Thus we have shown that  
the spin network quantum automaton recognizes `exactly'
(namely with a word--probability treshold $\delta =0$, see \eqref{tresh}
in section 2.1)
the braid group language according to the `Jones probability distribution'.\\
A similar result holds true for the families of automata $\mathcal{A}_{\,q}$
parametrized by $(j_1,j_2,$ $\ldots,j_{2n})$ that recognize
the braid group language $B_{2n}$ with a probability
distribution given by the colored polynomial
\eqref{colJon}.


\end{document}